\documentclass[zpreprint,zbstnp]{zeus_paper}

\usepackage[english]{babel}

\chardef\usc=95
\chardef\til=126
\catcode`\@=11 
\DeclareRobustCommand\xdotspace{\futurelet\@let@token\@xdotspace}
\def\@xdotspace{%
  \ifx\@let@token.\else
  \ifx\@let@token\bgroup.\else
  \ifx\@let@token\egroup.\else
  \ifx\@let@token\/.\else
  \ifx\@let@token\ .\else
  \ifx\@let@token~.\else
  \ifx\@let@token!.\else
  \ifx\@let@token,.\else
  \ifx\@let@token:.\else
  \ifx\@let@token;.\else
  \ifx\@let@token?.\else
  \ifx\@let@token/.\else
  \ifx\@let@token'.\else
  \ifx\@let@token).\else
  \ifx\@let@token-.\else
  \ifx\@let@token\@xobeysp.\else
  \ifx\@let@token\space.\else
  \ifx\@let@token\@sptoken.\else
   .\space
   \fi\fi\fi\fi\fi\fi\fi\fi\fi\fi\fi\fi\fi\fi\fi\fi\fi\fi}
\catcode`\@=12 

\newcommand{\stru}[2]{%
   \relax\ifmmode\hbox{\vrule height#1 depth#2 width0pt}%
   \else\vrule height#1 depth#2 width0pt\fi}

\newcommand{\Ronum}[1]{\uppercase\expandafter{\romannumeral#1}}
\newcommand{\ronum}[1]{\expandafter{\romannumeral#1}}
\DeclareRobustCommand{\LaTeXZ}{%
  \LaTeX\kern-.05em4\kern-.1em
  {\raisebox{-0.2ex}{$\scriptstyle\text{ZEUS}$}}\xspace}

\DeclareMathAlphabet{\mathbf}{OT1}{cmr}{bx}{sl}
\newcommand{\eVdist}{\kern-0.06667em}

\newcommand{\Gev}{{\text{Ge}\eVdist\text{V\/}}}

\newcommand{\pb}{\,\text{pb}}

\newcommand{\Tesla}{\,\text{T}}

\newcommand{\slashfrac}[2]{%
  \raisebox{0.5ex}{\ensuremath #1}\kern-0.12em/\kern-0.08em
  \raisebox{-.8ex}{\ensuremath #2}}

\newcommand{\sqr}[3]{%
    {\vcenter{\hrule height.#3ex\hbox{\vrule width.#2ex height#1ex
     \kern#1ex\vrule width.#3ex}\hrule height.#2ex}}}

\catcode`\@=11 
\newcommand{\parenbar}{\mathpalette\p@renb@r}
\def\p@renb@r#1#2{\vbox{%
  \ifx#1\scriptscriptstyle \dimen@.7em\dimen@ii.2em\else
  \ifx#1\scriptstyle \dimen@.8em\dimen@ii.25em\else
  \dimen@1em\dimen@ii.4em\fi\fi \offinterlineskip
  \ialign{\hfill##\hfill\cr
    \vbox{\hrule width\dimen@ii}\cr
    \noalign{\vskip-.3ex}%
    \hbox to\dimen@{$\mathchar300\hfil\mathchar301$}\cr
    \noalign{\vskip-.3ex}%
    $#1#2$\cr}}}
\catcode`\@=12 

\newcommand{\IP}{{\rm I$\kern-0.01667em$P}\xspace}

\mathchardef\qsm=63
\mathchardef\pls=43
\mathchardef\mns=512
\mathchardef\plm=518
\mathchardef\eql=61
\mathchardef\smallleft=300
\mathchardef\smallright=301
\mathchardef\les=316
\mathchardef\gre=318
\mathchardef\leq=532
\mathchardef\grq=533

\catcode`\@=11 
\newcounter{pict@width}
\newcounter{pict@height}
\newlength{\pict@scale}
\newcommand{\psfigadd}[4]{%
\setcounter{pict@width}{1*\ratio{#2+\pict@scale/2}{\pict@scale}}
\setcounter{pict@height}{1*\ratio{#3+\pict@scale/2}{\pict@scale}}
\setlength{\unitlength}{\pict@scale}
\hbox to #2{\hspace{-\fill}\begin{picture}(\thepict@width,\thepict@height)
\put(0,0){\psfig{figure=#1,width=#2,height=#3,clip=}}
\SetScale{0.283466457}
\SetWidth{1.763889}
{#4}
\end{picture}}
}
\newcounter{pict@widthfst}
\newcounter{pict@widthscd}
\newcounter{pict@widthtot}
\newcommand{\psfigaddtwo}[7]{%
\setcounter{pict@widthfst}{1*\ratio{#2+\pict@scale/2}{\pict@scale}}
\setcounter{pict@widthscd}{1*\ratio{#2+#4+\pict@scale/2}{\pict@scale}}
\setcounter{pict@widthtot}{1*\ratio{#2+#4+#6+\pict@scale/2}{\pict@scale}}
\setcounter{pict@height}{1*\ratio{#3+\pict@scale/2}{\pict@scale}}
\setlength{\unitlength}{\pict@scale}
\hbox{\hspace{-\fill}\begin{picture}(\thepict@widthtot,\thepict@height)
\put(0,0){\psfig{figure=#1,width=#2,height=#3,clip=}}
\put(\thepict@widthscd,0){\psfig{figure=#5,width=#6,height=#3,clip=}}
\SetScale{0.283466457}
\SetWidth{1.763889}
{#7}
\end{picture}}
}
\newcommand{\psfigror}[4]{%
\setcounter{pict@width}{1*\ratio{#2+\pict@scale/2}{\pict@scale}}
\setcounter{pict@height}{1*\ratio{#3+\pict@scale/2}{\pict@scale}}
\setlength{\unitlength}{\pict@scale}
\hbox{\begin{picture}(\thepict@width,\thepict@height)
\put(0,\thepict@height){\psfig{figure=#1,width=#3,height=#2,clip=,angle=270}}
\SetScale{0.283466457}
\SetWidth{1.763889}
{#4}
\end{picture}}
}
\newcommand{\psfigrol}[4]{%
\setcounter{pict@width}{1*\ratio{#2+\pict@scale/2}{\pict@scale}}
\setcounter{pict@height}{1*\ratio{#3+\pict@scale/2}{\pict@scale}}
\setlength{\unitlength}{\pict@scale}
\hbox{\begin{picture}(\thepict@width,\thepict@height)
\put(0,0){\psfig{figure=#1,width=#3,height=#2,clip=,angle=90}}
\SetScale{0.283466457}
\SetWidth{1.763889}
{#4}
\end{picture}}
}
\catcode`\@=12 
\newlength\listtextwidth

\catcode`\@=11 
\newlength{\@tabfninsert}
\newlength{\@tabfnwidth}
\newcommand{\tabfootnote}[2]{%
  \setlength{\@tabfninsert}{0.8em}
  \setlength{\@tabfnwidth}{\textwidth}
  \addtolength{\@tabfnwidth}{-\@tabfninsert}
  \addtolength{\@tabfnwidth}{-0.4em}
  \noindent\makebox[\@tabfninsert][r]{\footnotesize$^{#1}$\hfil}\hfill%
  \parbox[t]{\@tabfnwidth}{\footnotesize #2\hfill}}
\catcode`\@=12 

\def\JHEP{JHEP}

\def\etjet{E_T^{\rm jet}}

\def\etaphi{\eta-\varphi}

\def\setjb{d\sigma/d\etjb}
\def\sq2{d\sigma/d\q2}
 
\def\q2{Q^2}

\def\cgh{\cos\gamma_h}
\def\pb1{pb$^{-1}$}
\def\g2{GeV$^2$}

\def\kt{k_T}

\def\etjb{E^{\rm jet}_{T,{\rm B}}}
\def\etjbj{E^{\rm jet1}_{T,{\rm B}}}
\def\etjbjj{E^{\rm jet2}_{T,{\rm B}}}

\def\etajb{\eta^{\rm jet}_{\rm B}}
\def\etajbj{\eta^{\rm jet1}_{\rm B}}
\def\etajbjj{\eta^{\rm jet2}_{\rm B}}

\def\etalab{\eta^{\rm jet}_{\rm LAB}}
\def\etlab{E^{\rm jet}_{T,{\rm LAB}}}

\def\colab#1{#1 Coll.}

\def\z0{Z^0}
\def\mz{M_Z}

\def\as{\alpha_s}
\def\oalphas2{{\cal O}(\alpha\as^2)}

\def\oass{{\cal O}(\as^2)}

\def\asz{\as(\mz)}

\def\etal{et al.}

\def\qq{q\bar q}

\begin{document}

\prepnum{DESY--06--128}

\title{Inclusive-jet and dijet cross sections in deep inelastic
  scattering at HERA}

\author{ZEUS Collaboration}
\date{August 2006}

\abstract{
Inclusive-jet and dijet differential cross sections have been measured
in neutral current deep inelastic $ep$ scattering for 
exchanged boson virtualities $\q2>125$~\g2
with the ZEUS detector at HERA using an
integrated luminosity of $82$~\pb1. Jets were identified in the
Breit frame using the $\kt$ cluster algorithm.
Jet cross sections are presented as functions of several kinematic and jet
variables. The results are also presented in different regions of $\q2$. 
Next-to-leading-order QCD calculations
describe the measurements well. Regions of phase space where the
theoretical uncertainties are small have been identified. Measurements
in these regions have the potential to constrain the gluon density in the
proton when used as inputs to global fits of the proton parton
distribution functions.
}

\makezeustitle

\def\3{\ss}

\pagenumbering{Roman} 

\begin{center}                                                                                     
{                      \Large  The ZEUS Collaboration              }                               
\end{center}                                                                                       
  S.~Chekanov,                                                                                     
  M.~Derrick,                                                                                      
  S.~Magill,                                                                                       
  S.~Miglioranzi$^{   1}$,                                                                         
  B.~Musgrave,                                                                                     
  D.~Nicholass$^{   1}$,                                                                           
  \mbox{J.~Repond},                                                                                
  R.~Yoshida\\                                                                                     
 {\it Argonne National Laboratory, Argonne, Illinois 60439-4815}, USA~$^{n}$                       
\par \filbreak                                                                                     
  M.C.K.~Mattingly \\                                                                              
 {\it Andrews University, Berrien Springs, Michigan 49104-0380}, USA                               
\par \filbreak                                                                                     
  N.~Pavel~$^{\dagger}$, A.G.~Yag\"ues Molina \\                                                   
  {\it Institut f\"ur Physik der Humboldt-Universit\"at zu Berlin,                                 
           Berlin, Germany}
% ...
\newpage                                                                                           
$^{\    1}$ also affiliated with University College London, UK \\                                  
% ...
$^{\dagger}$ deceased \\

\newpage

\begin{tabular}[h]{rp{14cm}}                                                                       
$^{a}$ &  supported by the Natural Sciences and Engineering Research Council of Canada (NSERC) \\  
$^{b}$ &  supported by the German Federal Ministry for Education and Research (BMBF), under        
          contract numbers HZ1GUA 2, HZ1GUB 0, HZ1PDA 5, HZ1VFA 5\\                                
$^{c}$ &  supported in part by the MINERVA Gesellschaft f\"ur Forschung GmbH, the Israel Science   
          Foundation (grant no. 293/02-11.2) and the U.S.-Israel Binational Science Foundation \\  
$^{d}$ &  supported by the German-Israeli Foundation and the Israel Science Foundation\\           
$^{e}$ &  supported by the Italian National Institute for Nuclear Physics (INFN) \\                
$^{f}$ &  supported by the Japanese Ministry of Education, Culture, Sports, Science and Technology 
          (MEXT) and its grants for Scientific Research\\                                          
$^{g}$ &  supported by the Korean Ministry of Education and Korea Science and Engineering          
          Foundation\\                                                                             
$^{h}$ &  supported by the Netherlands Foundation for Research on Matter (FOM)\\                   
$^{i}$ &  supported by the Polish State Committee for Scientific Research, grant no.               
          620/E-77/SPB/DESY/P-03/DZ 117/2003-2005 and grant no. 1P03B07427/2004-2006\\             
$^{j}$ &  partially supported by the German Federal Ministry for Education and Research (BMBF)\\   
$^{k}$ &  supported by RF Presidential grant N 1685.2003.2 for the leading scientific schools and  
          by the Russian Ministry of Education and Science through its grant for Scientific        
          Research on High Energy Physics\\                                                        
$^{l}$ &  supported by the Spanish Ministry of Education and Science through funds provided by     
          CICYT\\                                                                                  
$^{m}$ &  supported by the Particle Physics and Astronomy Research Council, UK\\                   
$^{n}$ &  supported by the US Department of Energy\\                                               
$^{o}$ &  supported by the US National Science Foundation\\                                        
$^{p}$ &  supported by the Polish Ministry of Scientific Research and Information Technology,      
          grant no. 112/E-356/SPUB/DESY/P-03/DZ 116/2003-2005 and 1 P03B 065 27\\                  
$^{q}$ &  supported by FNRS and its associated funds (IISN and FRIA) and by an Inter-University    
          Attraction Poles Programme subsidised by the Belgian Federal Science Policy Office\\     
$^{r}$ &  supported by the Malaysian Ministry of Science, Technology and                           
Innovation/Akademi Sains Malaysia grant SAGA 66-02-03-0048\\                                       
\end{tabular}                                                                                      
\newpage

\pagenumbering{arabic} 
\pagestyle{plain}

\section{Introduction}

A well established testing ground for perturbative QCD (pQCD) is jet
production in neutral current (NC) deep inelastic $ep$ scattering
(DIS) at high $\q2$, where $\q2$ is the negative of the square of the
virtuality of the exchanged boson $V^{*}$ ($V=\gamma, Z$).
The parton-model process ($V^{*} q\rightarrow q$) in DIS gives rise to final states
containing one jet of high transverse energy, 
the so-called `current jet'. 
At leading order (LO) in $\as$, the boson-gluon-fusion 
($V^{*} g\rightarrow \qq$) and 
QCD-Compton-scattering ($V^{*} q \rightarrow qg$) processes 
give rise to two hard jets.

The hadronic final state in NC DIS may consist of jets with high
transverse energy, $\etjet$, produced in the hard-scattering process,
accompanied by the remnant (beam jet) of the incoming proton. 
For the analysis of these types of
process, the Breit frame~\cite{bookfeynam:1972,*zfp:c2:237} is
advantageous, since it provides a maximal separation between the products
of the beam fragmentation and the hard jets. In this frame, the exchanged
virtual boson is purely space-like, with 3-momentum ${\bf q}=(0,0,-Q)$, and is
collinear with the incoming parton. In the
parton-model process, the virtual boson $V^{*}$ is absorbed by the struck quark, which
is back-scattered with zero transverse momentum with respect to the $V^{*}$
direction, whereas the beam jet follows the direction of the initial struck
quark. 
QCD-Compton and boson-gluon-fusion events, on the other hand, are
characterised by two partons in the final state with possibly
non-vanishing transverse momentum with respect to the $V^{*}$
direction.
Thus, while retaining hard QCD processes at leading order in $\as$,
the contribution from the parton-model
process can be suppressed by requiring the production of jets with high
transverse energy in the Breit frame.
Therefore, measurements of inclusive-jet events at high
$\etjet$ or dijet events in the Breit frame are directly sensitive to
hard QCD processes, allowing direct tests of the pQCD predictions
and of the proton parton distribution functions (PDFs).

Jet cross sections in NC DIS have been measured previously at HERA. 
Inclusive-jet~\cite{pl:b547:164,epj:c19:289,pl:b542:193}, 
dijet~\cite{pl:b507:70} 
and multijet~\cite{epj:c44:183} production have been used to
extract values of $\as$. A previous dijet analysis~\cite{epj:c23:13}
has been used to test the gluon density extracted from global fits.
Results from 
inclusive-jet and dijet measurements~\cite{epj:c19:289} have been used
to constrain the gluon density in the proton. Recently, jet
cross sections in NC DIS~\cite{pl:b547:164} and
photoproduction~\cite{epj:c23:615} have been included in 
a new NLO QCD analysis to extract the proton PDFs~\cite{epj:c42:1}, 
resulting in a significant reduction of
the uncertainty on the gluon density at medium and high momentum fractions.

This paper presents new measurements of differential dijet cross
sections as functions of $\q2$, of the mean jet transverse energy of the
dijet system in the Breit frame, $\overline{E}_T$, of the dijet 
invariant mass, $M_{\rm jj}$, 
of the half-difference of the jet pseudorapidities
in the Breit frame, $\eta^{\prime}=|\etajbj-\etajbjj|/2$, and of the
fraction of the proton momentum taken by the interacting parton,
$\xi=x_{\rm Bj}(1+M_{\rm jj}^2/\q2)$. Here, $x_{\rm Bj}$ is the Bjorken
scaling variable that defines, for the parton-model process, the fraction of
proton momentum carried by the struck parton.
Measurements of the dijet cross section as a
function of $\xi$ are also shown for different regions of $\q2$. 
Also presented are measurements of the inclusive-jet cross section as a
function of the jet transverse energy in the Breit frame, $\etjb$, in
different regions of $\q2$. 
The analyses are based on data samples with more than twice the statistics of
the previous studies~\cite{pl:b547:164,epj:c23:13}.
These measurements probe an extended kinematic
regime with respect to previous analyses due to the increase in the proton
beam energy. 
The improvement in the experimental uncertainties obtained here will 
facilitate a more precise determination of the gluon density in the 
proton at high $\xi$.

\section{Experimental set-up}
A detailed description of the ZEUS detector can be found
elsewhere~\cite{pl:b293:465,zeus:1993:bluebook}. A brief outline of
the components that are most relevant for this analysis is given
below.

Charged particles are tracked in the central tracking detector
(CTD)~\cite{nim:a279:290,*npps:b32:181,*nim:a338:254}, which operates
in a magnetic field of $1.43\Tesla$ provided by a thin superconducting
solenoid. The CTD consists of $72$~cylindrical drift-chamber
layers, organized in nine superlayers covering the
polar-angle\footnote{The ZEUS coordinate system is a right-handed
  Cartesian system, with the $Z$ axis pointing in the proton beam
  direction, referred to as the ``forward direction'', and the $X$
  axis pointing left towards the centre of HERA. The coordinate origin
  is at the nominal interaction point. The polar angle, $\theta$, and the
  azimuthal angle, $\varphi$, are defined with respect to the proton beam
  direction. The pseudorapidity is defined as
  $\eta=-\ln\left(\tan\frac{\theta}{2}\right)$. }
region \mbox{$15^\circ<\theta<164^\circ$}. The transverse-momentum
resolution for full-length tracks can be parameterised as
$\sigma(p_T)/p_T=0.0058p_T\oplus0.0065\oplus0.0014/p_T$, with $p_T$ in
$\Gev$. The tracking system was used to measure the interaction vertex
with a typical resolution along (transverse to) the beam direction of
$0.4$~($0.1$)~cm and to cross-check the energy scale of the calorimeter.

The high-resolution uranium--scintillator calorimeter
(CAL)~\cite{nim:a309:77,*nim:a309:101,*nim:a321:356,*nim:a336:23} covers 
$99.7\%$ of the total solid angle and consists 
of three parts: the forward (FCAL), the barrel (BCAL) and the rear (RCAL)
calorimeters. 
Each part is segmented into towers and subdivided in depth into one
electromagnetic section and either one (in RCAL) or two (in FCAL and BCAL)
hadronic sections.
The smallest subdivision of the calorimeter is called a cell. Under
test-beam conditions, the CAL single-particle relative energy
resolutions were $\sigma(E)/E=0.18/\sqrt E$ for electrons and
$\sigma(E)/E=0.35/\sqrt E$ for hadrons, with $E$ in GeV.

The luminosity was measured from the rate of the bremsstrahlung process
$ep\rightarrow e\gamma p$. The resulting small-angle energetic photons
were measured by the luminosity
monitor~\cite{desy-92-066,*zfp:c63:391,*acpp:b32:2025}, a
lead-scintillator calorimeter placed in the HERA tunnel at $Z=-107$ m.

\section{Data selection and jet search}
The data used in this analysis 
were collected during the period 1998-2000, when HERA
operated with protons of energy $E_p=920$~GeV and electrons or
positrons\footnote{Here and in the following, the term ``electron''
  denotes generically both the electron and the positron.} 
of energy $E_e=27.5$~GeV, and
correspond to an integrated luminosity of $81.7\pm 1.8$~\pb1.

A three-level trigger system was used to select events
online~\cite{zeus:1993:bluebook,proc:chep:1992:222}. At the third
level, NC DIS events were accepted on the basis of the identification
of a scattered-electron candidate using localised energy depositions
in the CAL. 
An independent trigger selection which required at least one (two) jet(s) with
transverse energies above 10(6)~GeV and pseudorapidities below 2.5 was used to
measure the fraction of scattered electrons that gave a trigger signal. The
efficiency of the trigger selection based on the scattered-electron candidate
was found to be above 99$\%$.

Events were selected offline using criteria similar to those reported
previously~\cite{pl:b547:164}. The main steps are briefly listed
below. The inclusive-jet measurement is based on CAL cells for the
reconstruction of the kinematic and jet variables, whereas the dijet
analysis uses a combination of track and CAL 
information~\cite{phd:1998}. The selected combinations of 
tracks and CAL clusters are referred to as Energy Flow Objects (EFOs).
 
The scattered-electron candidate was identified from the pattern of
energy deposits in the CAL~\cite{nim:a365:508,*nim:a391:360}. The
energy ($E_e^{\prime}$) and polar angle ($\theta_e$) of the
electron candidate were determined from the CAL measurements. 
The kinematic variables $\q2$ and $x_{{\rm Bj}}$ were reconstructed 
using the double angle (DA)
method~\cite{proc:hera:1991:23,*proc:hera:1991:43}. This method uses
$\theta_e$ and the angle $\gamma_h$, which is equivalent to the angle of
the scattered quark in the quark-parton model. The angle $\gamma_h$, 
defined as

$$\cgh=\frac{(1-y)x_{\rm Bj}E_p-yE_e}{(1-y)x_{\rm Bj}E_p+yE_e},$$

was reconstructed using the hadronic 
final state~\cite{proc:hera:1991:23,*proc:hera:1991:43}, 
where $y=\q2/x_{\rm Bj}s$ and $s$ is the centre-of-mass energy.

The following requirements were imposed on the data sample for both the
inclusive-jet and dijet analyses:
\begin{itemize}
\item an electron candidate of energy $E_{e}^{\prime}>10$~GeV.
 This requirement ensured a high and well understood electron-finding
 efficiency and suppressed background from photoproduction events, in
 which the scattered electron escaped down the rear beampipe;
\item the total energy not associated with the electron candidate within
  a cone of radius 0.7 units in the pseudorapidity-azimuth ($\etaphi$)
  plane around the electron direction should be less than $10\%$ of
  the electron energy.
  This condition removed photoproduction and DIS events in which part
  of a jet was falsely identified as the scattered electron;
\item the vertex position along the beam axis should be in the range
  $|Z|<34$~cm. This condition helped to select events consistent with $ep$
  interactions;
\item $P_{T,{\rm miss}}/\sqrt{E_T}<2.5$~GeV$^{1/2}$, 
  where $P_{T,{\rm miss}}$ is the
  missing transverse momentum as measured with the CAL
  and $E_T$ is the total transverse energy in the CAL.
  This cut removed cosmic-ray events and beam-related background;
\item $\q2 > 125$~\g2; 
\item $-0.65 < \cgh < 0.65$. 
  The lower limit avoided a region with limited acceptance due to the
  requirement on the energy of the scattered electron, while the
  upper limit was chosen to ensure good reconstruction of the jets in
  the Breit frame.
\end{itemize}

In addition, the following selection requirements were imposed for the
inclusive-jet analysis:
\begin{itemize}
\item $y_e<0.95$, where $y_e=1-E_{e}^{\prime}(1-\cos{\theta_{e}})/(2 E_e)$.
  This condition removed events in which fake electron candidates were
  found in the FCAL;
\item if the polar angle 
  of the scattered electron was in the range
  $30^{\circ}<\theta_e<140^{\circ}$, 
  it was required that the fraction of the
  electron energy within a cone of radius 0.3 units in the $\etaphi$
  plane around the electron direction should be larger than 0.9; for
  $\theta_e<30^{\circ}$, the cut was raised to 0.98. This condition removed
  events in which jets were misidentified as electrons;
\item the event was not consistent with elastic Compton scattering  
  ($ep\rightarrow e\gamma p$), namely
  no second electromagnetic energy cluster above 10~GeV was allowed when the rest 
  of the CAL energy, besides the two electromagnetic energy clusters, was below
  4~GeV; 
\item $38<(E-P_Z)<65$~GeV, where $E$ is the total energy,
  $E=\sum_iE_i$, and $P_Z$ is the $Z$-component of the vector 
  ${\bf P}=\sum_i \bf{p_i}$. The sums run over all final-state objects.
  This cut removed events with large initial-state radiation and
  further reduced the background from photoproduction.      
\end{itemize}

For the dijet analysis, the following conditions had to be fulfilled: 
\begin{itemize}
\item $\q2 < 5000$~\g2;
\item $45<(E-P_Z)<62$~GeV. 
\end{itemize}

The $\kt$ cluster algorithm~\cite{np:b406:187} was used in the
longitudinally invariant inclusive mode~\cite{pr:d48:3160} to
reconstruct jets in the hadronic final state both in data and in Monte
Carlo (MC) simulated events (see Section~\ref{mc}) assuming massless
objects. In data, the 
algorithm was applied to the final-state objects after excluding
the scattered-electron candidate. The jet search
was performed in the $\etaphi$ plane of the Breit frame. The jet
variables were defined according to the Snowmass
convention~\cite{proc:snowmass:1990:134}.

After reconstructing the jet variables in the Breit frame, the
massless four-momenta were boosted into the laboratory frame, where
the transverse energy~($\etlab$) and the pseudorapidity~($\etalab$) of
each jet were calculated. Energy
corrections~\cite{pl:b547:164,epj:c23:615,pl:b531:9,pl:b558:41} were then applied
to the jets in the laboratory frame and propagated into $\etjb$ 
in order to compensate for
energy losses in the inactive material in front of the CAL. 
The following cuts were applied: 
\begin{itemize}
\item events were removed from the sample if the distance $\Delta$ of any of
  the jets to the electron candidate in the $\etaphi$ plane of the
  laboratory frame was smaller than 1~unit,
  $\Delta=\sqrt{(\etalab-\eta^e)^2+(\varphi_{\rm LAB}^{\rm jet}-\varphi^e)^2}<1$,
  where $\varphi^e$ and $\eta^e$ are the azimuthal angle and pseudorapidity of
  the scattered electron, respectively.
  This requirement removed some background from photoproduction and
  improved the purity of the sample;
\item events were removed from the sample if a jet was in
  the backward region of the detector ($\etalab<-2$).
  This requirement removed events in which a radiated photon from the
  electron was misidentified as a jet in the Breit frame;
\item $-2<\etajb<1.5$. This cut restricted the jets to a region with large
  acceptance. 
\end{itemize}

The inclusive-jet sample was then selected using the following conditions:
\begin{itemize}
\item $\etlab>2.5$~GeV.
  This cut removed a small number of jets for which the uncertainty on
  the energy correction was large;
\item $\etjb>8$~GeV.
\end{itemize}

For the dijet analysis, the following further requirements were imposed: 
\begin{itemize}
\item $\etlab>3$~GeV; 
\item $\etalab<2.5$. This condition restricted the jets to the region in
  which the reconstruction of the jet variables was optimal;
\item the two highest-$\etjb$ jets in an event, ordered according to
$\etjb$, were required to satisfy $\etjbj>12$~GeV and
$\etjbjj>8$~GeV. 
\end{itemize}
The selected sample of inclusive-jet (dijet) events consisted of  19908 (3868) events.

\section{Monte Carlo simulation and acceptance corrections}
\label{mc}
Samples of MC events were generated to determine the response of the
detector to jets of hadrons and the correction factors necessary to
obtain the hadron-level jet cross sections. 
The hadron level is defined in terms of hadrons with lifetime $\tau\geq 10$~ps. 
The generated events were
passed through the {\sc Geant}~3.13-based~\cite{tech:cern-dd-ee-84-1} ZEUS
detector- and trigger-simulation
programs~\cite{zeus:1993:bluebook}. They were reconstructed and
analysed by the same program chain as the data.
 
Neutral current DIS events including radiative effects were simulated
using the {\sc Heracles}~4.6.1~\cite{cpc:69:155,*spi:www:heracles}
program with the {\sc Djangoh}~1.1~\cite{cpc:81:381,*spi:www:djangoh11} 
interface to the hadronisation programs. {\sc Heracles} includes
corrections for initial- and final-state radiation, vertex and
propagator terms, and two-boson exchange. The QCD cascade is simulated
using the colour-dipole model
(CDM)~\cite{pl:b165:147,*pl:b175:453,*np:b306:746,*zfp:c43:625}
including the LO QCD diagrams as implemented in 
{\sc Ariadne}~4.08~\cite{cpc:71:15,*zfp:c65:285} and, alternatively,
with the MEPS model of {\sc Lepto}~6.5~\cite{cpc:101:108}. The
CTEQ5D~\cite{epj:c12:375} proton PDFs were used for these
simulations. Fragmentation into hadrons is performed using the Lund
string model~\cite{prep:97:31} as implemented in 
{\sc Jetset}~7.4~\cite{cpc:82:74,*cpc:39:347,*cpc:43:367}.
 
The jet search was performed on the MC events using the energy
measured in the CAL cells or EFOs in the same way as for the
data. 
The same jet algorithm was also
applied to the final-state particles (hadron level) and to the partons
available after the parton shower (parton level). 

\begin{figure}[p]
\vfill
\centerline{\epsfig{figure=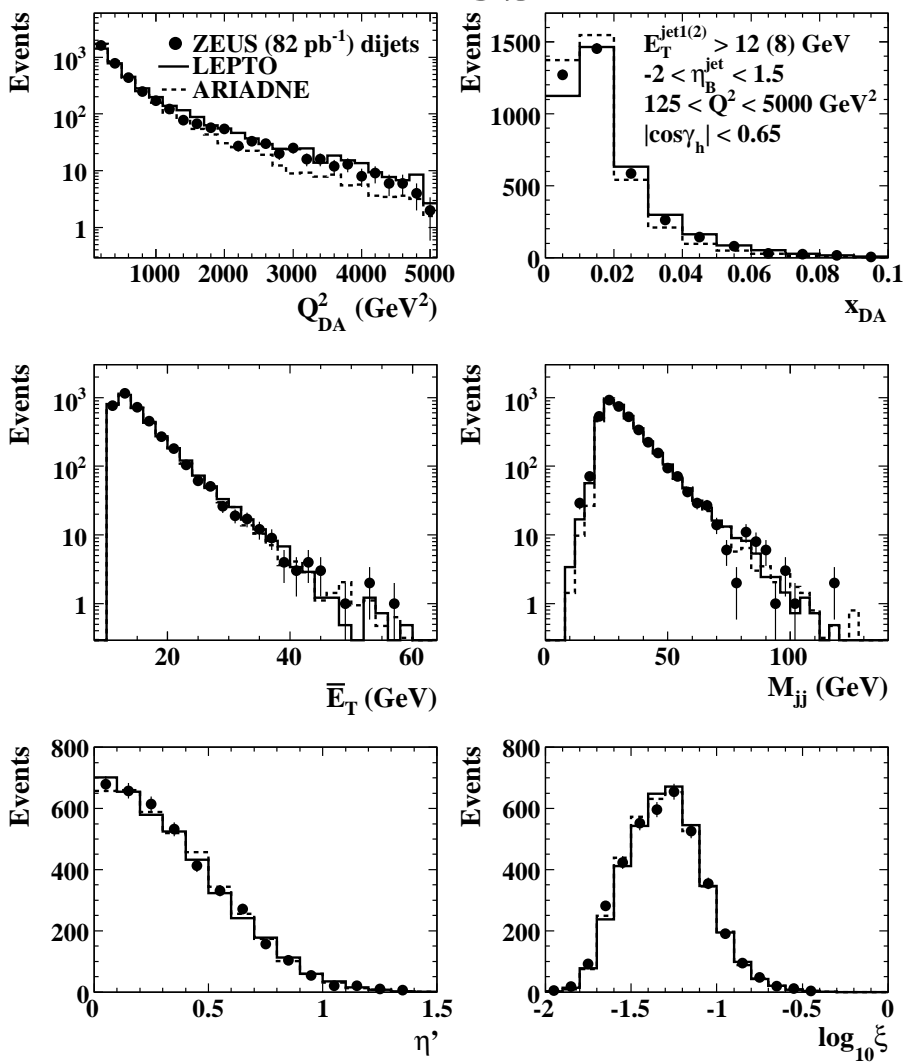,width=12.0cm}}
\vspace{-.5cm}
\caption
{\it 
Uncorrected data distributions for dijet production 
with $\etjbj>12$~{\rm GeV}, $\etjbjj>8$~{\rm GeV} and $-2<\etajb<1.5$
in the kinematic range given by $|{\rm cos} \gamma_h| <$~0.65 and 125~$< Q^2
<$~5000~${\rm GeV^2}$ (dots). For comparison, the predictions of the ARIADNE
(dashed histograms) and LEPTO (solid histograms) MC models are
also included. 
}
\label{mccontrol}
\vfill
\end{figure}

The data were corrected to the
hadron level and for QED-radiative effects using bin-by-bin correction factors
obtained from the MC samples. 
For this approach to be valid, the uncorrected distributions 
of the data must be well described by the MC simulations.
This condition was in general satisfied by both the {\sc Ariadne} and
{\sc Lepto} MC. 
The {\sc Ariadne} model gave a slightly better 
description of the inclusive-jet data and was thus used as the default
model; {\sc Lepto} 
was then used to estimate the systematic effect on the
correction procedure due to the parton-shower model (see Section~\ref{expunc}).
In contrast to this, the dijet distributions, 
shown in Fig.~\ref{mccontrol}, were
slightly better described by the {\sc Lepto} model; here, {\sc Ariadne}
was therefore used for systematic checks.
In all cases, the  correction factors differed from unity by typically 
$10\%$. These correction factors took into account the efficiency of the
trigger, the selection criteria and the purity and efficiency of the
jet reconstruction.

\section{NLO QCD calculations}
\label{nlo}
The measurements were compared with NLO QCD ($\oass$) calculations
obtained using the program {\sc Disent}~\cite{np:b485:291}. The
calculations were performed in the $\overline{\rm MS}$ renormalisation and
factorisation schemes using a generalised version~\cite{np:b485:291}
of the subtraction method~\cite{np:b178:421}. The number of flavours
was set to five and the factorisation scale was chosen to be
$\mu_F=Q$. Calculations with different choices of the renormalisation
scale, $\mu_R$, were performed: the default choice 
was $\mu_R^2=(E^{\rm jet}_{T,{\rm B}})^2$ for the
inclusive-jet analysis; cross checks were performed using $\mu_R^2=\q2$ and
$\mu_R^2=\q2+\overline{E}_{T,{\rm B}}^2$, 
where $\overline{E}_{T,{\rm B}}^2$ is the mean
transverse energy of the selected jets in an event. For the dijet analysis
 $\mu_R^2=\q2+\overline{E}_T^2$ was used as default, 
since the resulting predictions 
describe the data better than either scale alone; here the effects of using
either $\q2$ or $\overline{E}_T^2$ were investigated. 
The effects on the NLO calculations of using different choices for $\mu_R$
were found to be smaller than the uncertainties described below.
The strong coupling constant was calculated at two loops with
$\Lambda^{(5)}_{\overline{\rm MS}}=226$~MeV, corresponding to
$\asz=0.118$. The calculations were performed using the 
CTEQ6~\cite{jhep:0207:012,*jhep:0310:046} parameterisations of the
proton PDFs. The $\kt$ cluster algorithm was also applied to the
partons in the events generated by {\sc Disent} in order to obtain
the jet cross-section predictions. 

Since the measurements refer to jets of hadrons, whereas the NLO QCD
calculations refer to jets of partons, the predictions were corrected
to the hadron level using the MC models. The multiplicative correction
factor ($C_{\rm had}$) was defined as the ratio of the cross section
for jets of hadrons over that for jets of partons, estimated by using 
the MC programs described in Section~\ref{mc}. The ratios obtained
with {\sc Ariadne} and {\sc Lepto} were averaged to obtain the value of
$C_{\rm had}$. This value differs from unity by approximately 
$5\ (5-10)\%$ for the inclusive-jet (dijet) analysis.

The NLO QCD predictions for the inclusive-jet analysis were also
corrected for the $\z0$-exchange contribution by using MC simulated
events with and without $\z0$ exchange. This correction is not
required for the dijet analysis owing to the restricted $\q2$ range.

Several sources of uncertainty in the theoretical predictions were
considered:
\begin{itemize}
  \item the uncertainty on the NLO QCD calculations due to terms
    beyond NLO, estimated by varying $\mu_R$ by a factor of two up and down,
    was below $\pm 7\%$ at low $\q2$ and low $\etjb$ and
    decreased to below $\pm 5\%$ in the highest-$\q2$ region for the 
    inclusive-jet analysis.
    For the dijet measurement, the average uncertainty was about $\pm
    10\%$, with the largest values approaching $\pm 20\%$ at low $\xi$ and low
    $Q^2$ values. 
  \item the uncertainty on the NLO QCD calculations due to that on
    $\asz$ was estimated by repeating the calculations using two
    additional sets of proton PDFs, CTEQ6A114 and
    CTEQ6A122~\cite{jhep:0602:032}, determined assuming $\asz=0.114$
    and $0.122$, respectively. The difference between the calculations
    using these sets and CTEQ6 was scaled by a factor of $0.68$ to
    reflect the uncertainty on
    $\as$~\cite{jp:g26:r27}.
    The resulting uncertainty in the
    cross sections was typically below $\pm 4\%$;
  \item the uncertainty on the modelling of the parton shower was
    estimated as half the difference between the multiplicative correction
    factors calculated from the {\sc Lepto} and {\sc Ariadne} models.
    The resulting uncertainty on the cross sections was typically less than
    $3\%$; 
  \item the uncertainty on the NLO calculations due to the
    proton PDFs was estimated by repeating the
    calculations using 40 additional sets from the CTEQ6 analysis, which takes
    into account the 
    statistical and correlated systematic experimental uncertainties of each
    data set used in the determination of the proton PDFs. 
    The resulting
    uncertainty in the cross sections was about $\pm 4\%$ at low $\q2$
    and decreased to around $\pm 2\%$ at high $\q2$;
  \item the uncertainty of the calculations in the value of $\mu_F$ was
    estimated by repeating the calculations with $\mu_F=Q/2$ and $2Q$.
    The variation of the calculations was negligible.
\end{itemize}
 
\begin{figure}[p]
\vfill
\setlength{\unitlength}{1.0cm}
\begin{picture} (18.0,15.0)
\put (1.0,0.0){\epsfig{figure=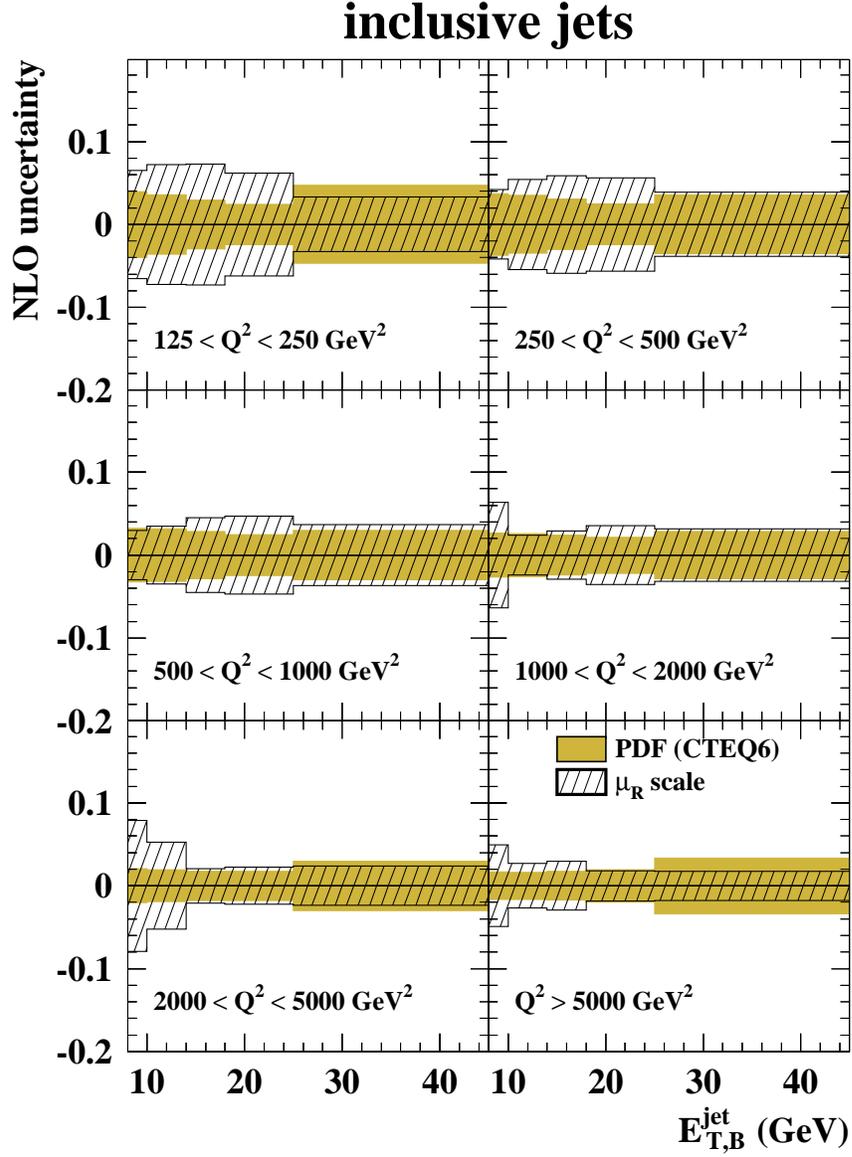,width=16cm}}
\end{picture}
\caption
{\it
Overview of theoretical uncertainties  
for the inclusive-jet analysis as functions of $\etjb$ in bins of $Q^2$
for $|{\rm cos} \gamma_h| <$~0.65 and $-2<\etajb<1.5$. Shown are the
relative uncertainties induced by the variation of the renormalisation scale
$\mu_R$ (hatched area) and by the uncertainties on the proton PDFs (shaded
area).
}
\label{fig11}
\vfill
\end{figure}

\begin{figure}[p]
\vfill
\setlength{\unitlength}{1.0cm}
\begin{picture} (18.0,15.0)
\put (1.0,0.0){\epsfig{figure=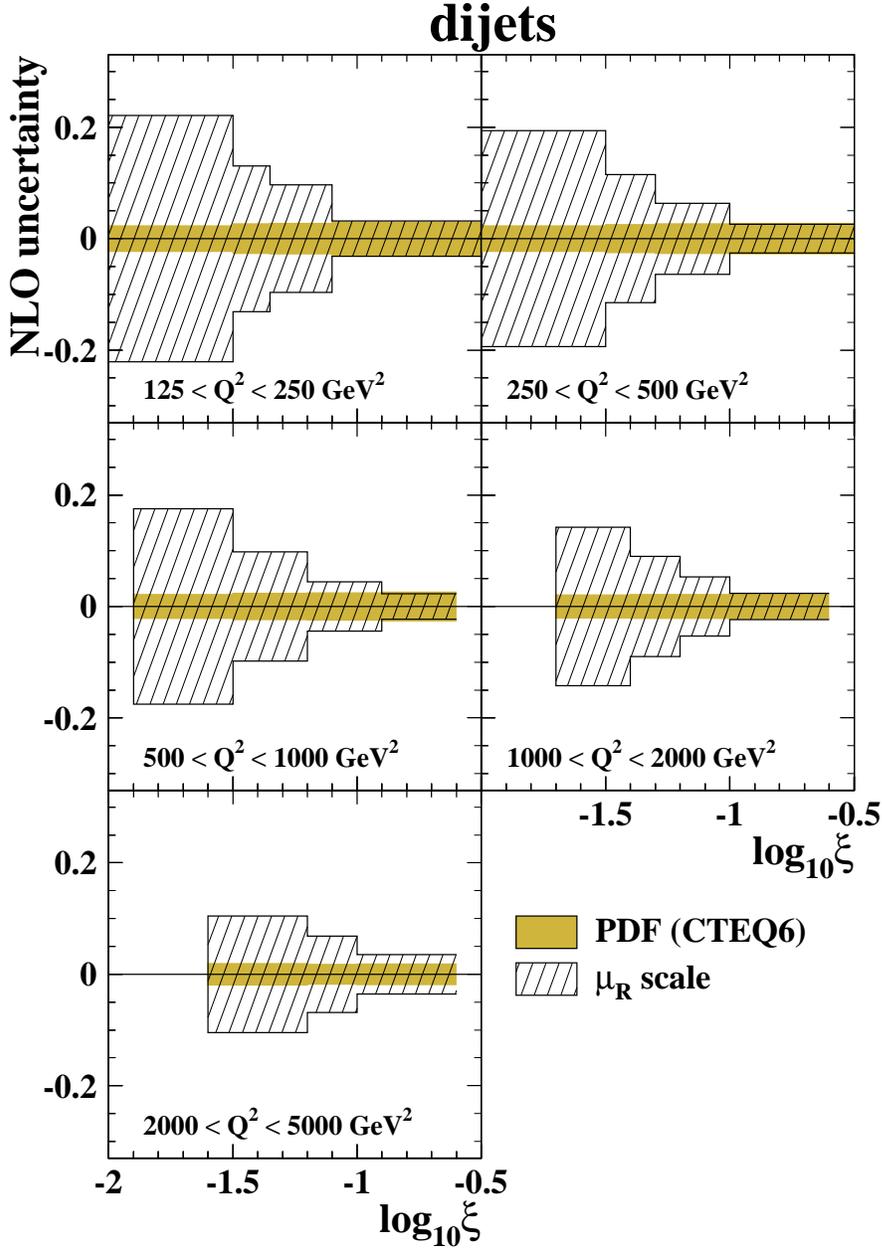,width=13cm}}
\end{picture}
\caption
{\it
Overview of theoretical uncertainties  
for the dijet analysis as functions of $\log_{10}\xi$ in 
bins of $Q^2$ for $|{\rm
  cos} \gamma_h| <$~0.65, $E_{T,{\rm B}}^{jet1} >$~12~{\rm GeV}, $E_{T,{\rm
    B}}^{jet2} >$~8~{\rm GeV} and $-2<\etajb<1.5$. Other details as in the
caption to Fig.~\ref{fig11}.}
\label{fig11a}
\vfill
\end{figure}

The total theoretical uncertainty was obtained by adding in quadrature
the individual uncertainties listed above.
Figures~\ref{fig11} and~\ref{fig11a} show an
overview of various theoretical uncertainties for both the inclusive-jet and
the dijet analysis. 

\begin{figure}[p]
\vfill
\setlength{\unitlength}{1.0cm}
\begin{picture} (18.0,15.0)
\put (1.0,0.0){\epsfig{figure=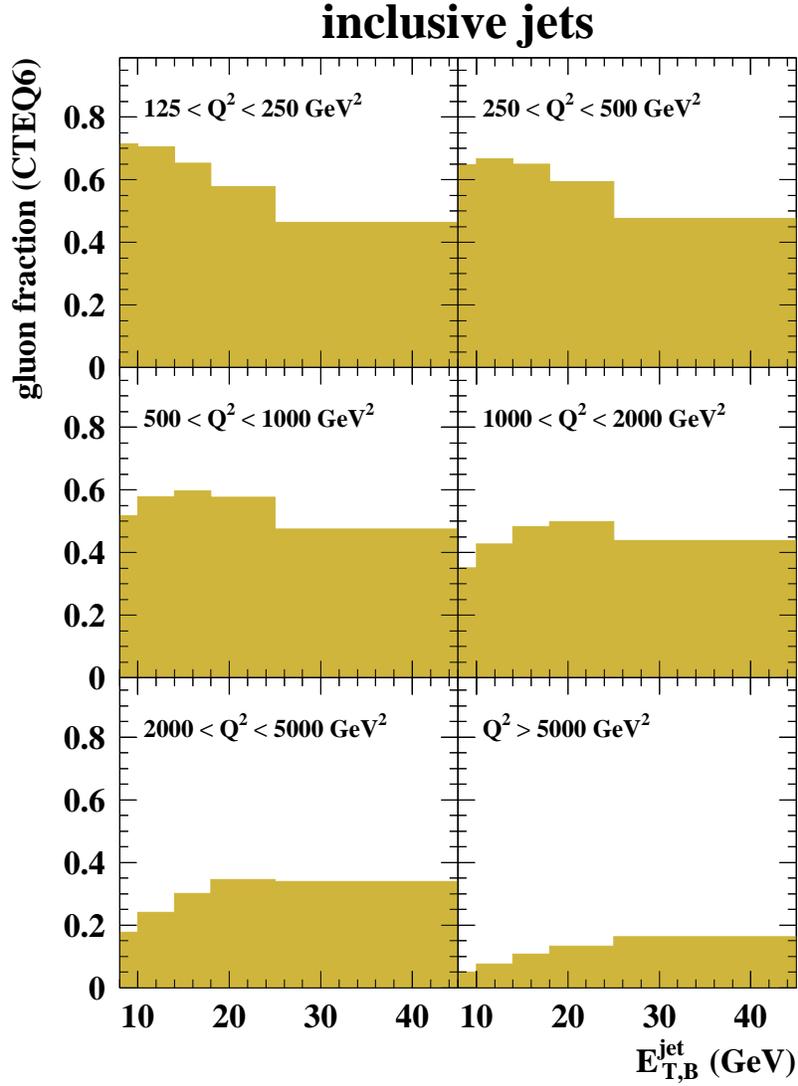,width=15cm}}
\end{picture}
\caption
{\it NLO QCD predictions of the gluon-induced fraction
  of the inclusive-jet cross sections $d\sigma/d\etjb$ as
  functions of $E_{T,B}^{jet}$ in different regions of $Q^2$ for $|{\rm cos}
  \gamma_h| <$~0.65 and $-2<\etajb<1.5$ (shaded histograms). The CTEQ6 
  parameterisations of the proton PDFs were used.}  
\label{gluefrac1}
\vfill
\end{figure}

\begin{figure}[p]
\vfill
\setlength{\unitlength}{1.0cm}
\begin{picture} (18.0,15.0)
\put (3.0,-1.0){\epsfig{figure=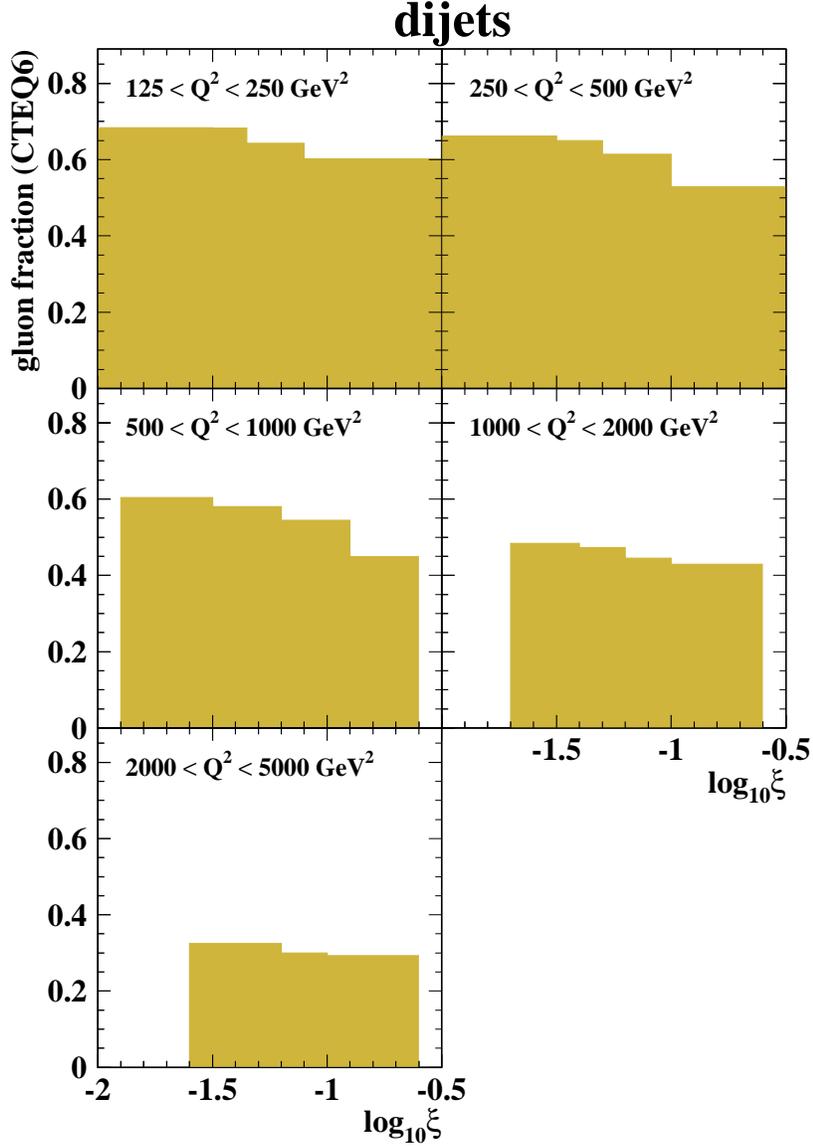,width=12cm}}
\end{picture}
\caption
{\it NLO QCD predictions of the gluon-induced fraction
  of the dijet cross-sections $d\sigma/d\log_{10}\xi$ as
  functions of $\log_{10}\xi$ in different regions of $Q^2$ for $|{\rm cos}
  \gamma_h| <$~0.65, $E_{T,{\rm B}}^{jet1} >$~12~{\rm GeV}, $E_{T,{\rm
    B}}^{jet2} >$~8~{\rm GeV} and $-2<\etajb<1.5$. Other details as in the
  caption to Fig.~\ref{gluefrac1}.}
\label{gluefrac2}
\vfill
\end{figure}

\begin{figure}[p]
\vfill
\setlength{\unitlength}{1.0cm}
\begin{picture} (18.0,15.0)
\put (1.0,0.0){\epsfig{figure=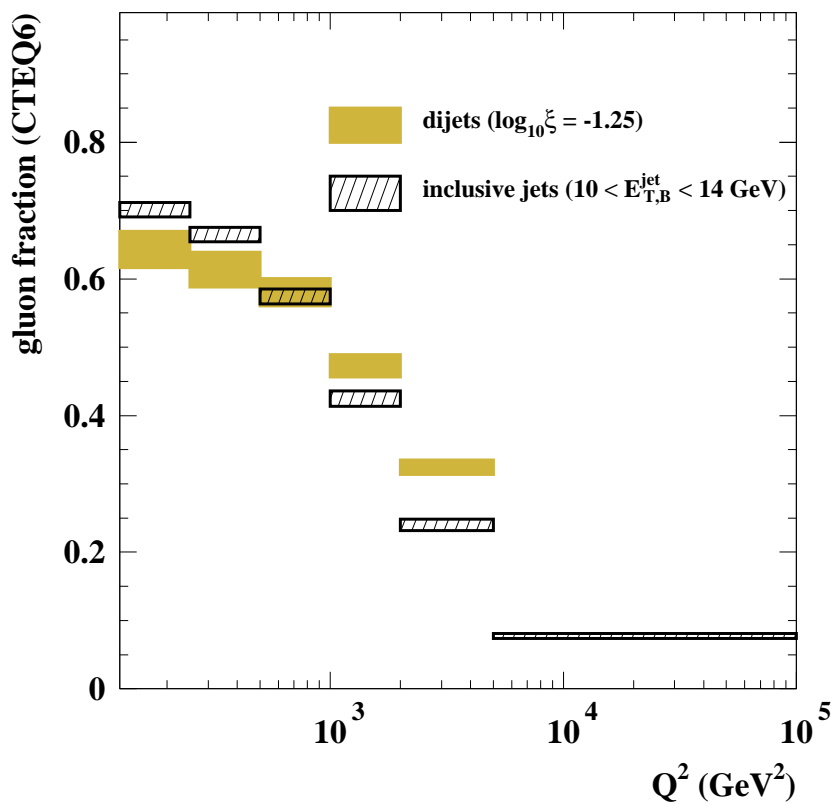,width=15cm}}
\end{picture}
\caption
{\it NLO QCD predictions of the gluon-induced fraction of the dijet
  (inclusive-jet) cross-section $d\sigma/d\log\xi$ ($d\sigma/d\etjb$) as a
  function of $Q^2$  for $|{\rm cos}
  \gamma_h| <$~0.65 and $-2<\etajb<1.5$ and for \mbox{$\log_{10}\xi =$~--1.25}
  \mbox{(10~$< \etjb <$~14~{\rm GeV})}; the 
  uncertainty on the gluon-induced fraction due to those on the proton PDFs is
  represented by the shaded (hatched) area. The CTEQ6 parameterisations of the
  proton PDFs were used. }
\label{gluefrac}
\vfill
\end{figure}

The gluon-induced contribution to the cross section as a function of $\etjb$
for the inclusive-jet analysis and of $\log_{10}\xi$ for the dijet analysis in
various regions of $\q2$ is shown in Figs.~~\ref{gluefrac1}
and~\ref{gluefrac2}. 
The gluon fraction varies between 70$\%$ at low $Q^2$ for both
low $\xi$ and low $\etjb$ and less than 10$\%$ for the lowest $\etjb$ at high
values of $Q^2$. For the dijet analysis, the gluon contribution is always at
least 30$\%$. In addition, Fig.~\ref{gluefrac} shows, separately for
the inclusive-jet and dijet analyses, the gluon-induced contribution to the
cross section as a function of $\q2$ together with the uncertainties as taken
from the CTEQ6 analysis.

\section{Experimental uncertainties}
\label{expunc}
The following sources of systematic uncertainty were considered for
the measured cross sections:
\begin{itemize}
  \item the uncertainty in the absolute energy scale of the jets was
    estimated to be $\pm 1\%$ for $\etlab>10$~GeV and $\pm 3\%$ for
    lower $\etlab$
    values~\cite{pl:b531:9,epj:c23:615,proc:calor:2002:767}. The
    resulting uncertainty on the cross sections
    was about $\pm 5\%$ and increased to 
    approximately $\pm 10\%$ in certain regions of the dijet phase space;
  \item the uncertainty in the absolute energy scale of the electron
    candidate was estimated to be $\pm 1\%$~\cite{epj:c21:443}. The
    resulting uncertainty was below $\pm 1\%$;
  \item the differences in the results obtained by using either
    {\sc Ariadne} or {\sc Lepto} to correct the data for detector
    effects were typically below $\pm8\%$;
  \item the analysis was repeated using an alternative
    technique~\cite{epj:c11:427} to select the scattered-electron
    candidate. The resulting uncertainty was typically below $\pm 3\%$;
  \item the $\etlab$ cut was raised to 4(4.5)~GeV for the inclusive-jet
    (dijet) analysis. The resulting uncertainty was
    typically smaller than $\pm 1\%$;
  \item the cut in $\etalab$ used to suppress the contamination due to
    photons falsely identified as jets in the Breit frame was changed to
    $-3$ and to $-1.5$. The resulting uncertainty was typically below 
    $\pm 1\%$;
  \item the uncertainty due to the selection cuts was estimated by
    varying the values of the cuts within the resolution of each
    variable. The effect on the cross sections was typically below
    $\pm 2\%$;
  \item the uncertainty on the cross sections due to that in the
    simulation of the trigger was negligible.
\end{itemize}

The systematic uncertainties not associated with the absolute energy
scale of the jets were added in quadrature to the statistical uncertainties
and are shown in figures~\ref{fig110} to~\ref{fig9} as error bars. 
The uncertainty due to the absolute energy scale of the jets is shown 
separately as a shaded band in each of these figures,
due to the large bin-to-bin correlation. In addition, there was an
overall normalisation uncertainty of $2.2\%$ from the luminosity
determination, which is not included in the figures.

\section{Results}
\label{results}

\subsection{Dijet differential cross sections}

Dijet cross sections were measured in the kinematic region
$125<\q2<5000$~\g2\ and $|\cgh|<0.65$. These cross sections correspond
to the two highest-$\etjb$ jets in each event with $\etjbj>12$~GeV,
$\etjbjj>8$ GeV and $-2<\etajb<1.5$ and were corrected for detector
and QED radiative effects as described in Section~\ref{mc}. 

\begin{figure}[p]
\vfill
\setlength{\unitlength}{1.0cm}
\begin{picture} (18.0,15.0)
\put (1.0,0.0){\epsfig{figure=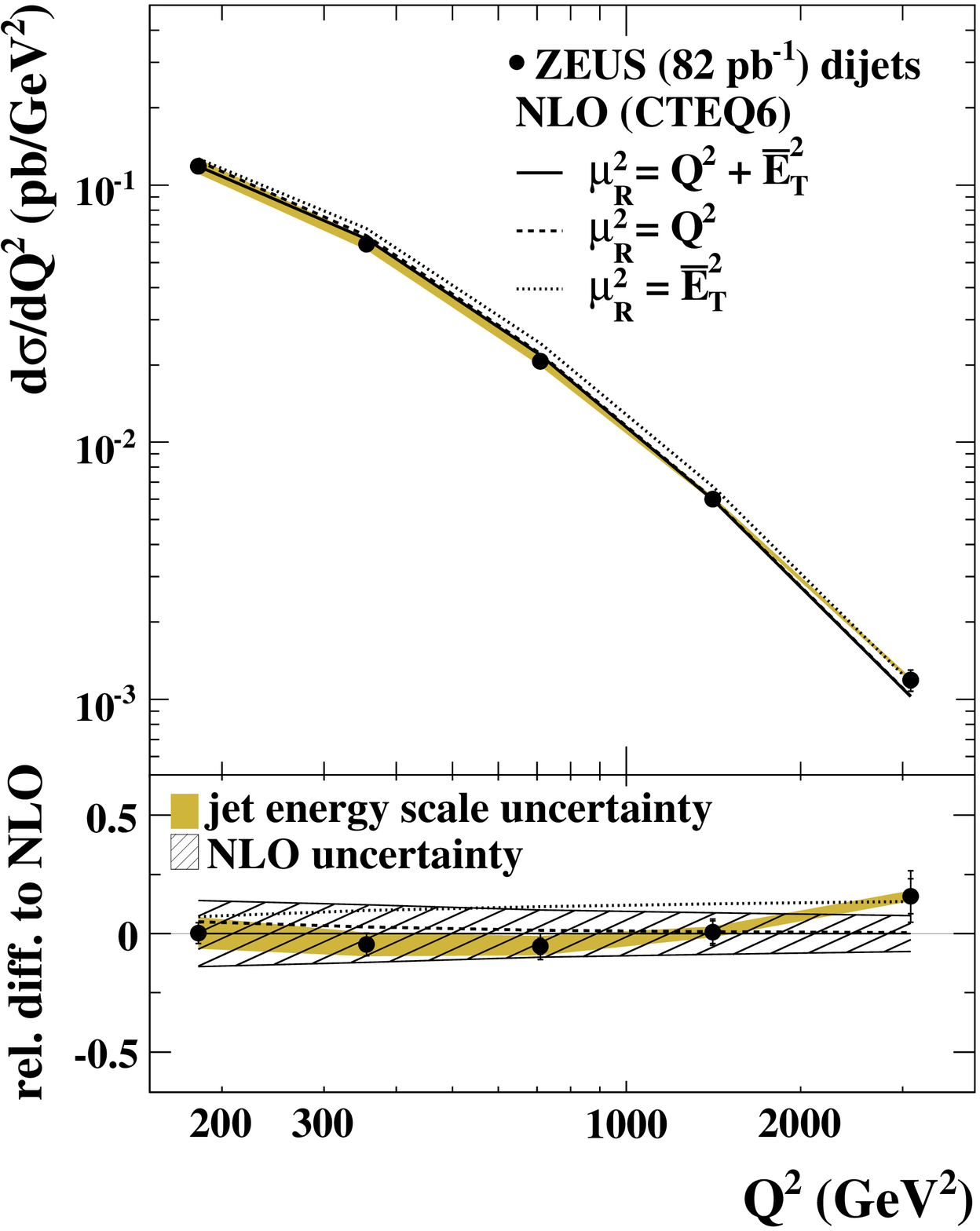,width=7.5cm}}
\put (8.5,0.0){\epsfig{figure=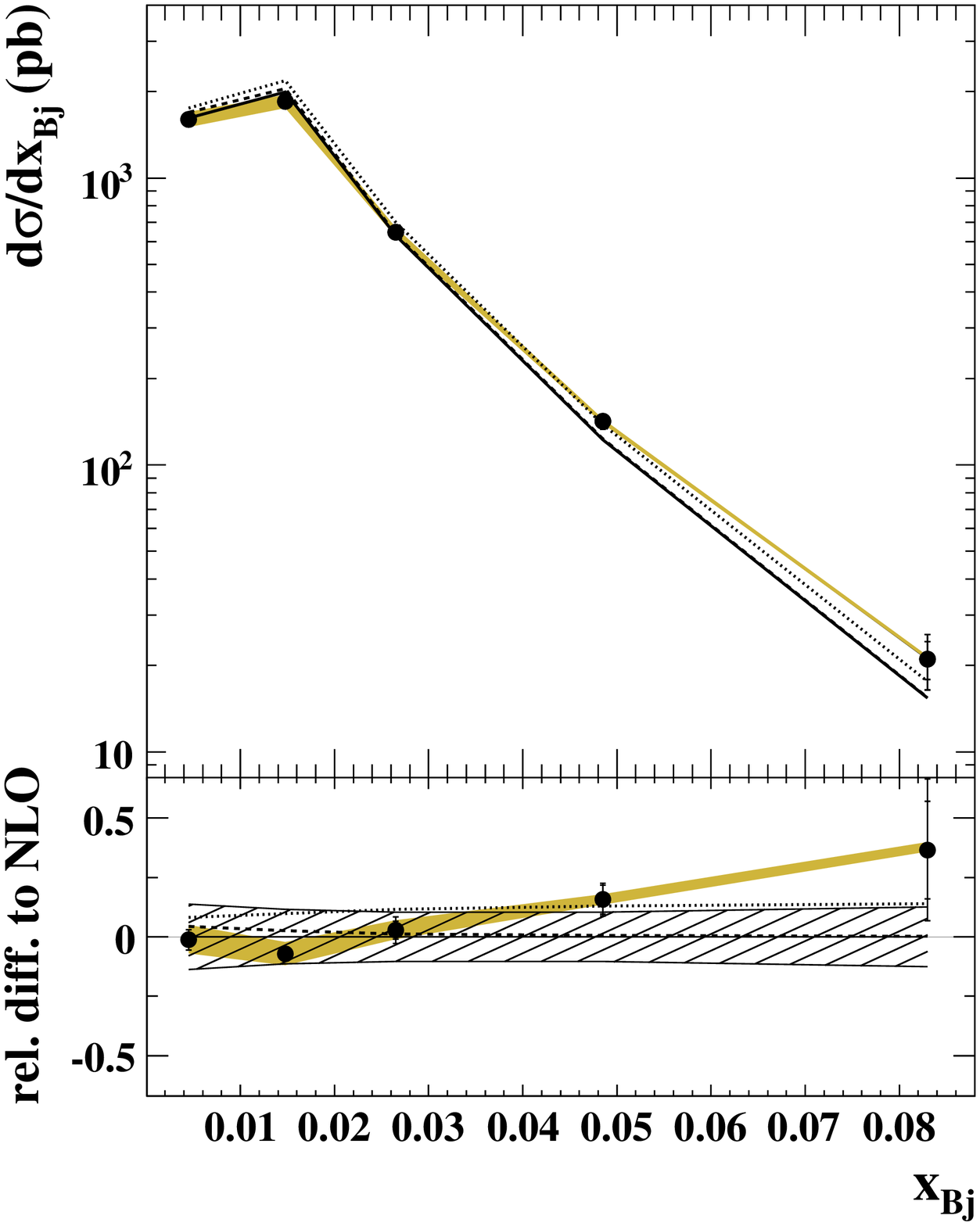,width=7.5cm}}
\put (7.25,9.0){{\bf {\Huge ZEUS}}}
\put (2.5,3.9){{\bf (a)}}
\put (10.,3.9){{\bf (b)}}
\end{picture}
\caption
{\it The measured differential cross-sections (a) $\sq2$ and (b)
  $d\sigma/dx_{{\rm Bj}}$
for dijet production with $\etjbj>12$~{\rm GeV}, $\etjbjj>8$~{\rm GeV} and $-2<\etajb<1.5$
(dots), in the kinematic range given by $|{\rm cos} \gamma_h| <$~0.65 and 125~$< Q^2
<$~5000~${\rm GeV^2}$. The inner error bars represent the statistical
uncertainty. The outer error bars show the statistical and systematic
uncertainties, not associated with the uncertainty in the absolute energy
scale of the jets, added in quadrature. The shaded bands display the
uncertainties due to the absolute energy scale of the jets. The NLO QCD
calculations with $\mu_R^2=\q2+\overline{E}_T^2$ (solid lines), $\mu_R^2=\q2$
(dashed lines) and \mbox{$\mu_R^2=\overline{E}_T^2$} (dash-dotted lines), corrected
for hadronisation effects and using the CTEQ6 parameterisations of the
proton PDFs, are also shown. The lower parts of the figures show the
relative differences with respect to the NLO QCD calculations with
\mbox{$\mu_R^2=\q2+\overline{E}_T^2$}: measurements (dots), NLO QCD calculations with
$\mu_R^2=\q2$ (dashed lines) and with $\mu_R^2=\overline{E}_T^2$ (dotted lines); 
the hatched bands display the total theoretical uncertainty. }
\label{fig110}
\vfill
\end{figure}

\begin{figure}[p]
\vfill
\setlength{\unitlength}{1.0cm}
\begin{picture} (18.0,15.0)
\put (1.0,8.5){\epsfig{figure=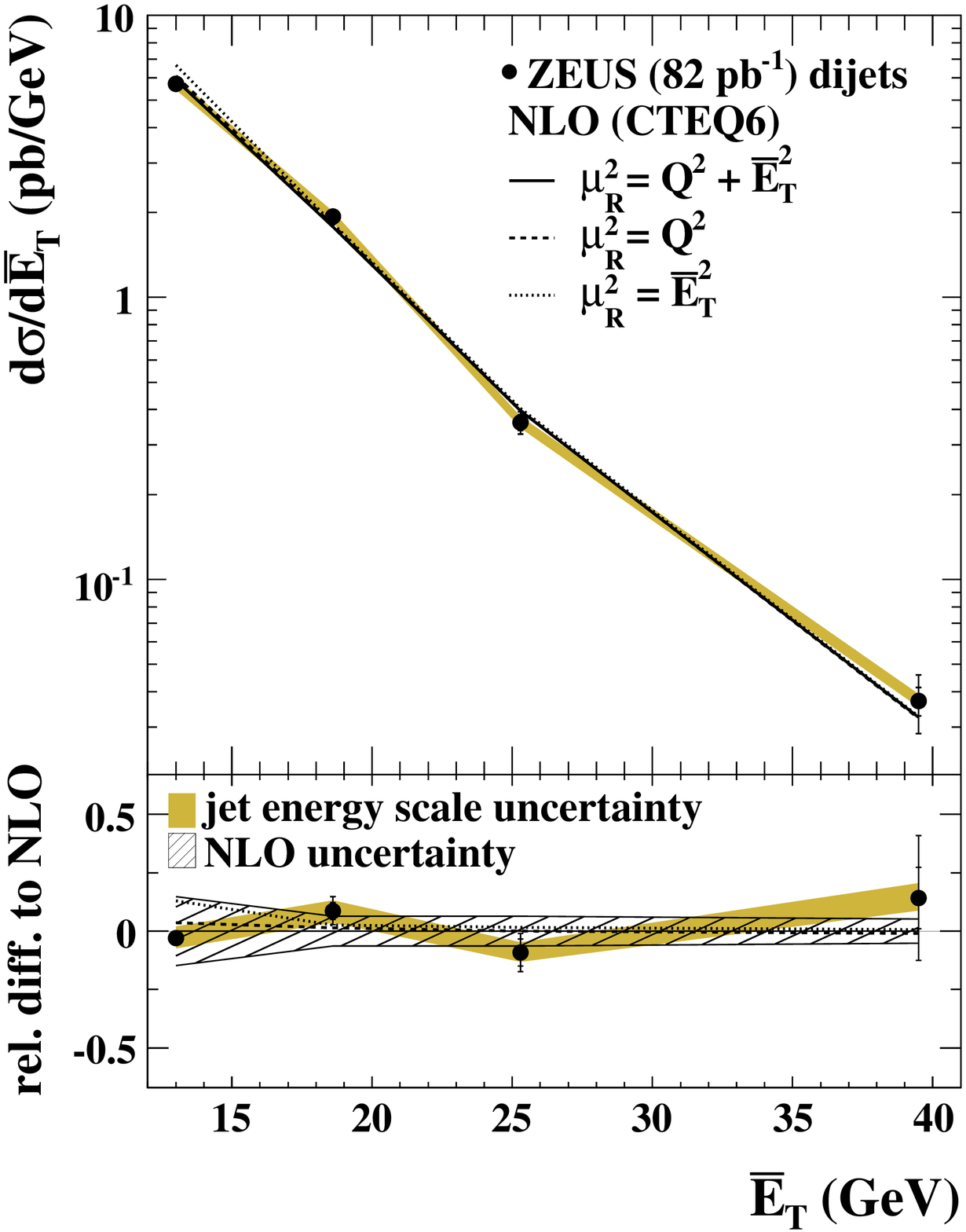,width=7.5cm}}
\put (8.5,8.5){\epsfig{figure=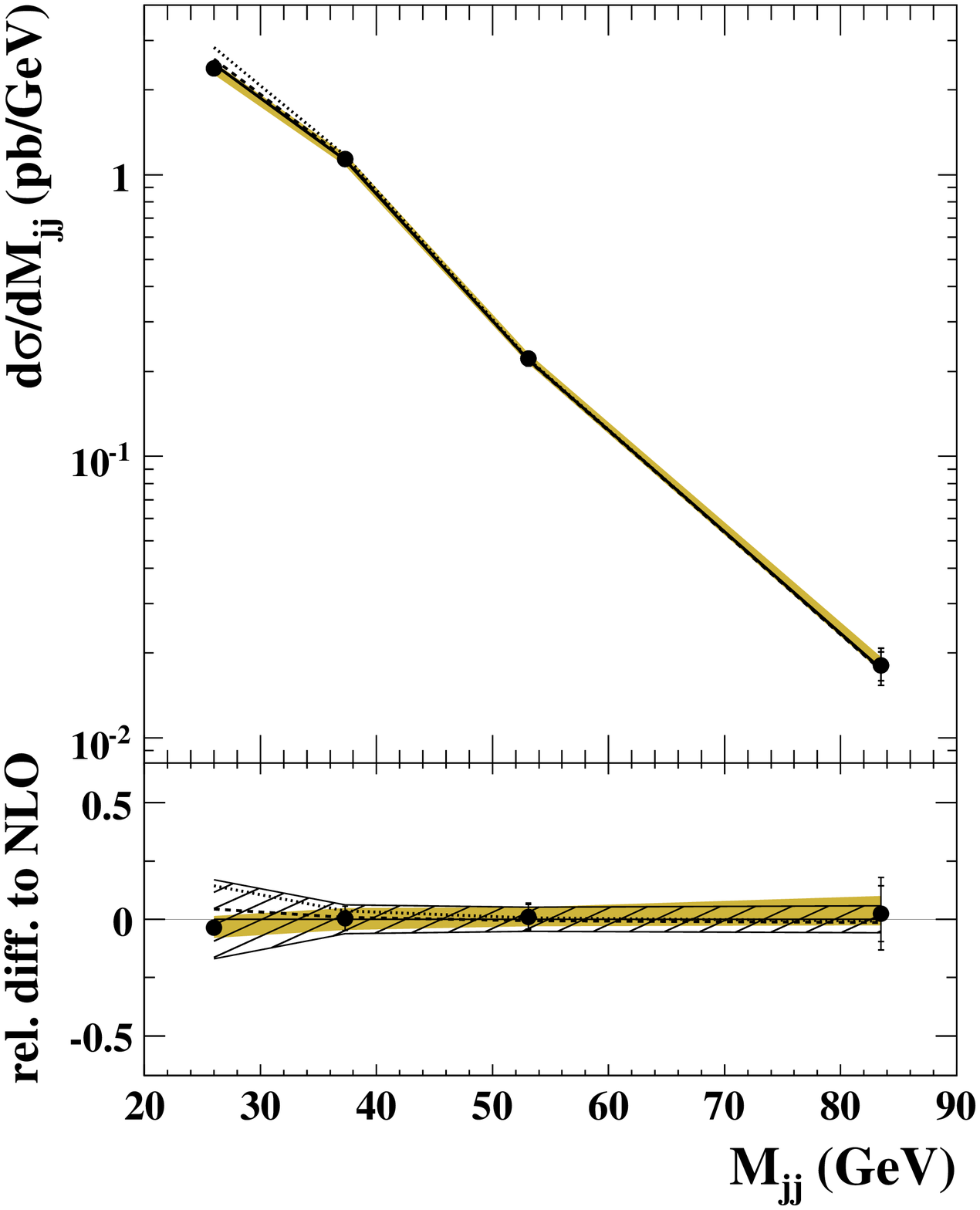,width=7.5cm}}
\put (1.0,-1.){\epsfig{figure=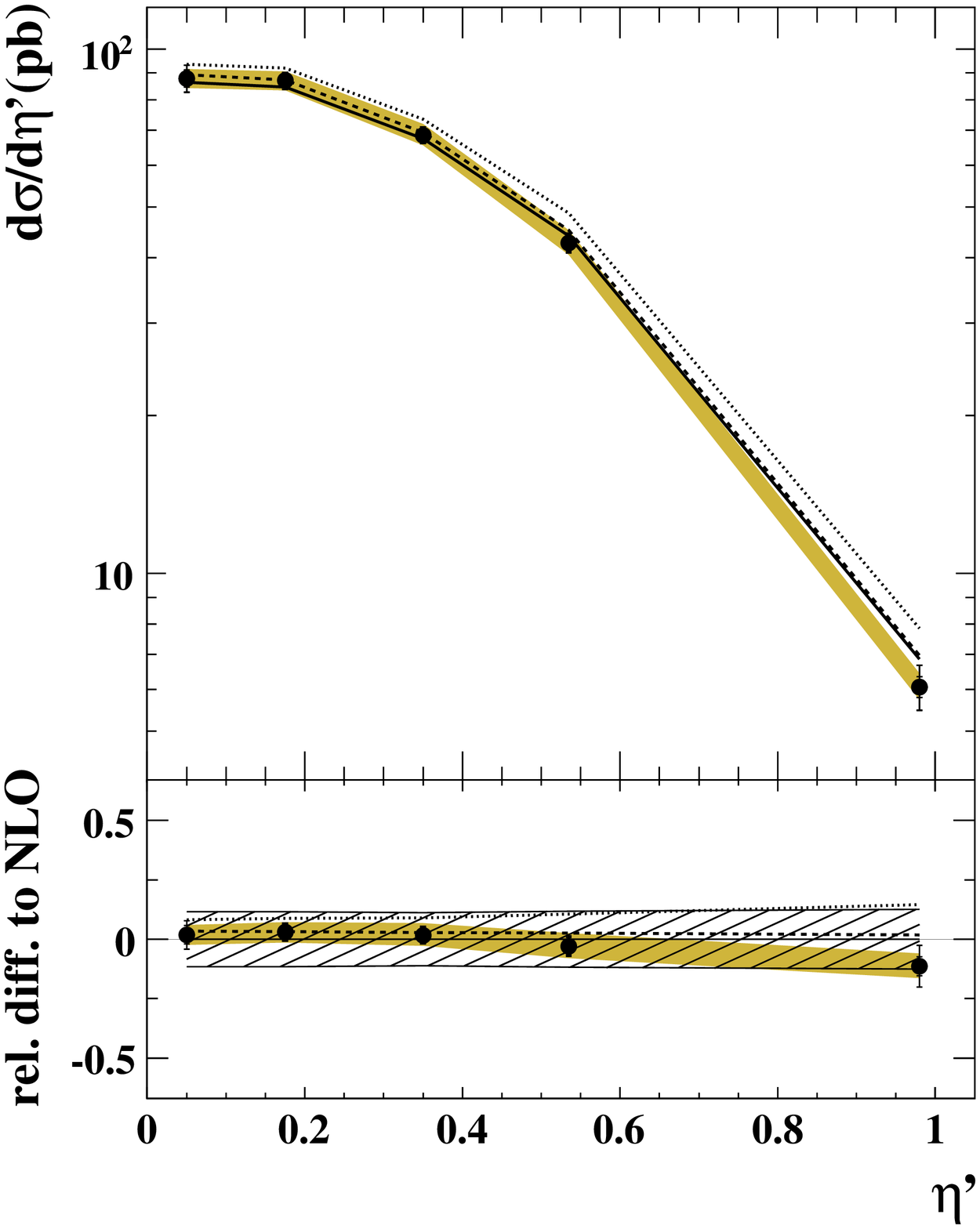,width=7.5cm}}
\put (8.5,-1.){\epsfig{figure=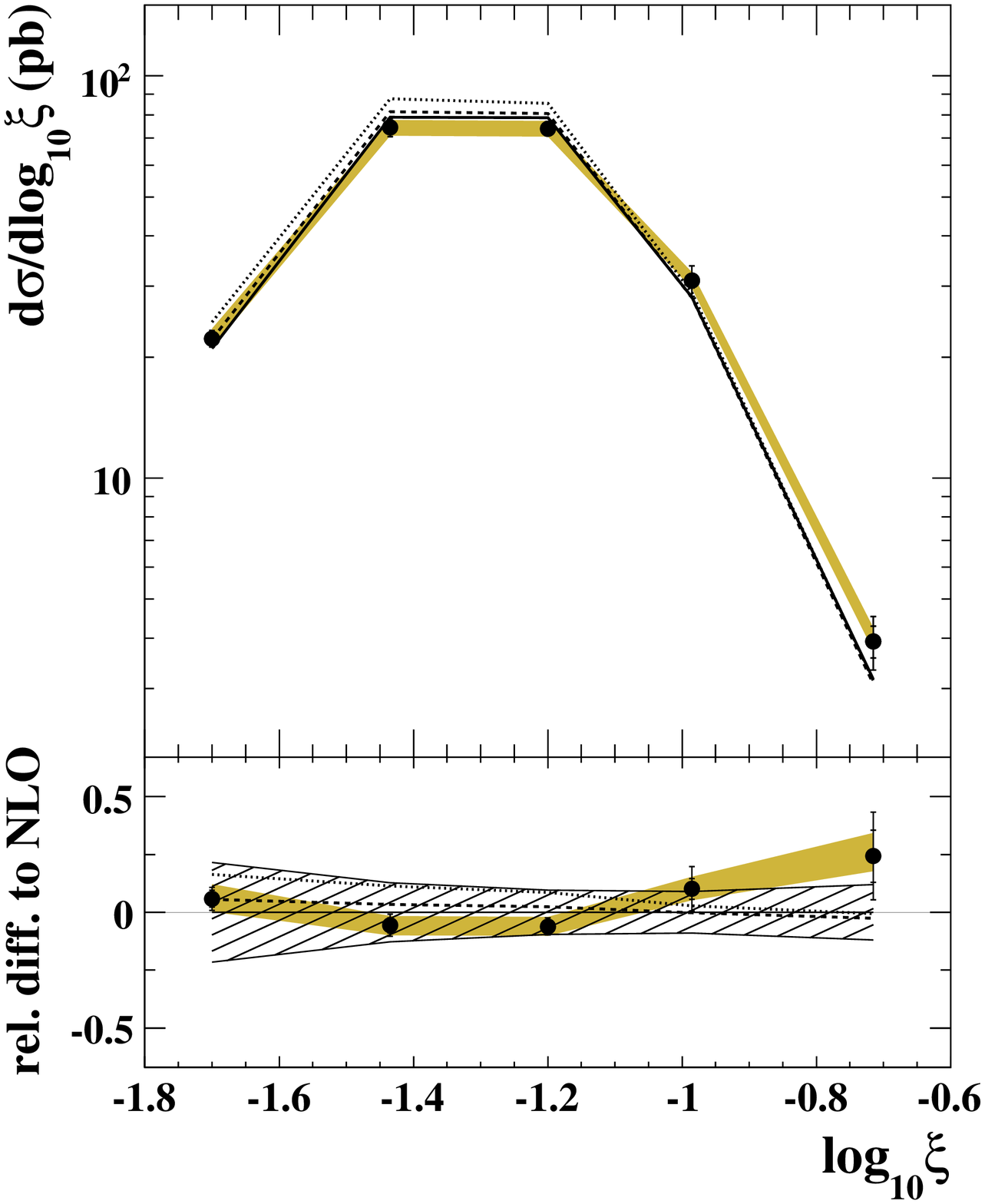,width=7.5cm}}
\put (7.25,17.5){{\bf {\Huge ZEUS}}}
\put (2.5,12.4){{\bf (a)}}
\put (10.,12.4){{\bf (b)}}
\put (2.5,2.9){{\bf (c)}}
\put (10.,2.9){{\bf (d)}}
\end{picture}
\vspace{1.cm}
\caption
{\it 
The measured differential cross-sections (a) $d\sigma/d\overline{E}_T$,
(b) $d\sigma/dM_{\rm jj}$, (c) $d\sigma/d\eta'$ and (d) $d\sigma/d\log_{10}\xi$
for dijet production
with $\etjbj>12$~{\rm GeV}, $\etjbjj>8$~{\rm GeV} and \mbox{$-2<\etajb<1.5$} (dots), in
the kinematic range given by $|{\rm cos} \gamma_h| <$~0.65 and 
\mbox{125~$< Q^2 <$~5000~${\rm GeV^2}$}.  Other details as in the caption to
Fig.~\ref{fig110}. }
\label{fig111}
\vfill
\end{figure}

\begin{figure}[p]
\vfill
\centerline{\epsfig{figure=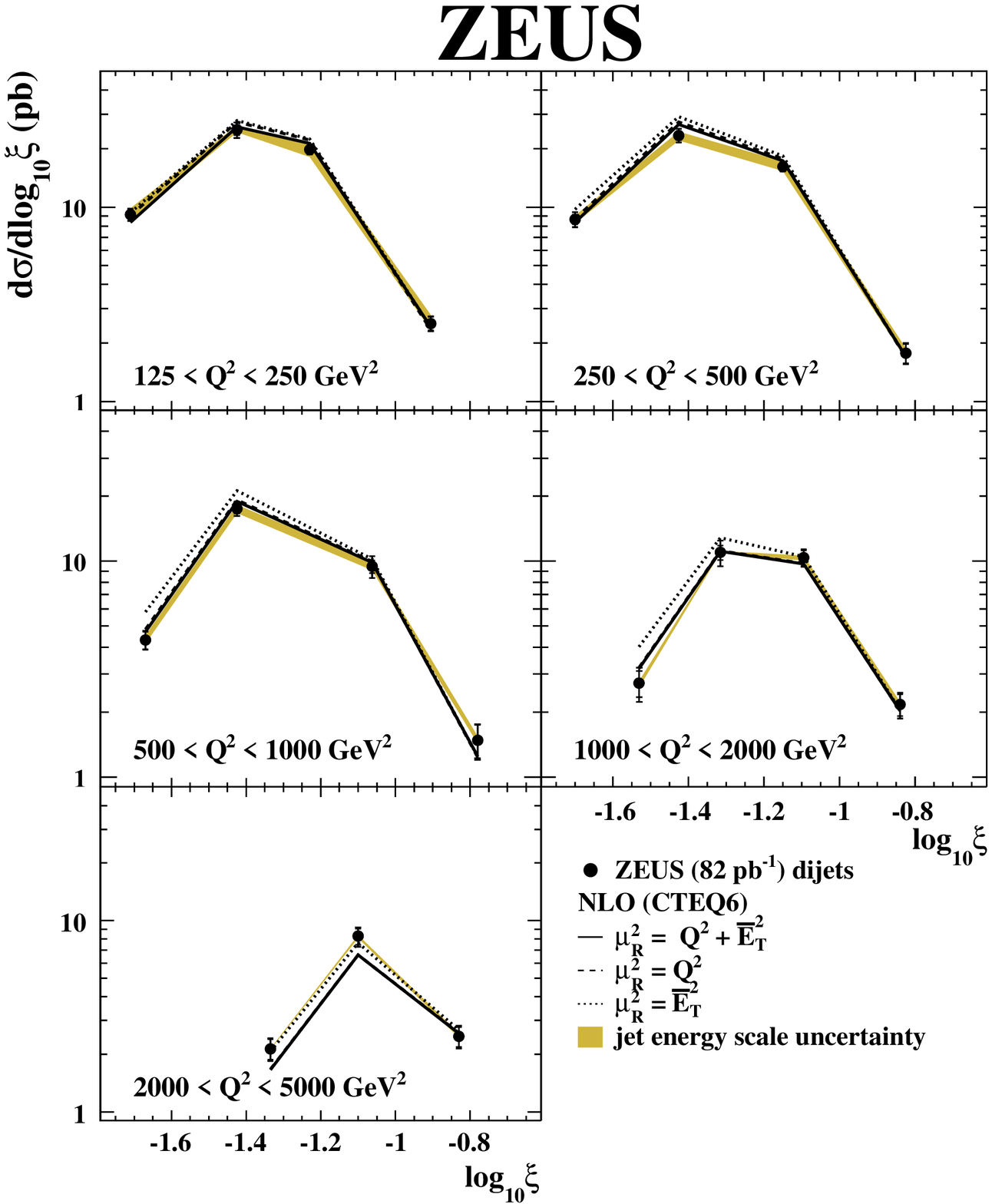,width=15.0cm}}
\caption
{\it
The measured differential cross-section $d\sigma/d\log_{10}\xi$ for
dijet production with $\etjbj>12$~{\rm GeV}, $\etjbjj>8$~{\rm GeV} and
$-2<\etajb<1.5$ in different regions of $\q2$ (dots). Other
details as in the caption to Fig.~\ref{fig110}.
}
\label{fig6}
\vfill
\end{figure}

\begin{figure}[p]
\vfill
\centerline{\epsfig{figure=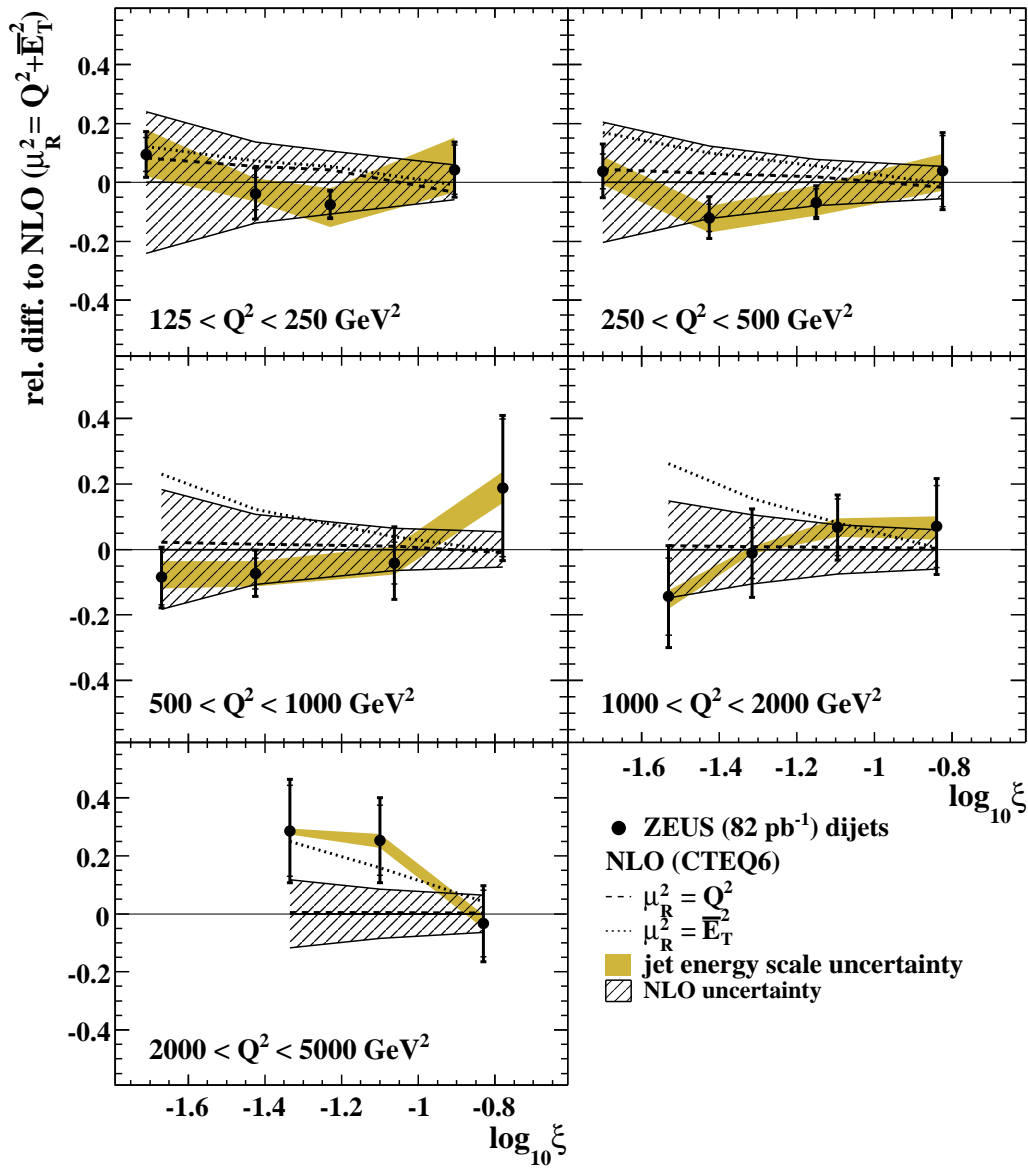,width=15.0cm}}
\caption
{\it
Relative differences between the measured differential cross-sections
$d\sigma/d\log_{10}\xi$ presented in Fig.~\ref{fig6} and the NLO QCD
calculations with $\mu_R^2=\q2+\overline{E}_T^2$ (dots). The relative
differences between the NLO QCD calculations with $\mu_R^2=\q2$ 
($\mu_R^2=\overline{E}_T^2$) and those with $\mu_R^2=\q2+\overline{E}_T^2$ are
also shown as dashed (dotted) lines.
Other details as in the caption to Fig.~\ref{fig110}.
}
\label{fig7}
\vfill
\end{figure}

The measured cross sections as functions of $\q2$, $x_{\rm Bj}$,
$\overline{E}_T$, $M_{\rm jj}$, $\eta^{\prime}$ 
and $\log_{10}\xi$ are shown in
Figs.~\ref{fig110} and~\ref{fig111} and are 
listed in Tables~\ref{q2_inc_dijet_tab} to~\ref{xi_inc_dijet_tab}. 
In these and subsequent figures, each data point is plotted at the abscissa at
which the differential cross section was equal to its bin-averaged value,
according to the NLO QCD calculation. 
The data distribution as a function of
$\q2$ exhibits a fall-off of more than two orders of magnitude
within the range studied. The $\q2$-restricted range implicitly limits
the $\xi$ range of the measurements to
$0.009\lesssim \xi \lesssim 0.37$, a region of phase space where
these measurements can significantly constrain the gluon density in
the proton. The dijet cross sections as functions of $\overline{E}_T$
and $M_{\rm jj}$ are particularly suited to test the matrix elements in the
perturbative calculations. The measured distributions show a steep
fall-off of more than two orders of magnitude within the measured range.

The NLO QCD calculations with $\mu_R^2=\q2+\overline{E}_T^2$ 
are compared to the data in Figs.~\ref{fig110} 
and~\ref{fig111}. They give a good description of the data.
The calculations with
$\mu_R^2=\q2$ or $\mu_R^2=\overline{E}_T^2$ 
are included in Figs.~\ref{fig110} and \ref{fig111} as well. 
At low $\q2$, low $M_{\rm jj}$ and low $\overline{E}_T$, the
theoretical uncertainty is dominated by the effect of the variation of
the renormalisation scale.
In the case of the dijet cross
sections as functions of $\overline{E}_T$ and $M_{\rm jj}$, the theoretical
uncertainty is smaller than the experimental at high values of 
$\overline{E}_T$ and $M_{\rm jj}$. The excellent
agreement between data and theory for these distributions demonstrates
the validity of the description of the dynamics of dijet production by
pQCD at $\oalphas2$.

\subsection{Dijet and inclusive-jet differential cross sections in
  different regions of $\q2$}

Figure~\ref{fig6} shows the measured dijet cross section as a function of 
$\log_{10}\xi$ in different regions of $\q2$. 
The cross sections are also given in
Table~\ref{dd_xi_inc_dijet_tab}. The requirement that two
jets with high transverse energy be observed in the final state
suppresses the cross section in the low-$\xi$ region, so that the
measured cross section rises as $\log_{10}\xi$ increases.
At high values of $\log_{10}\xi$, the decrease of the cross section reflects 
the decrease of the gluon and quark densities at high $\xi$.

The NLO QCD predictions with $\mu_R^2=\q2+\overline{E}_T^2$ are
compared to the measurements in Fig.~\ref{fig7} using the relative
difference of the measured differential cross sections to the NLO
calculations; the uncertainty of the calculation is also shown in the figure.  
The predictions of NLO QCD give a good
description of the data. Also indicated are the predictions using 
$\mu_R^2=\q2$ or $\mu_R^2=\overline{E}_T^2$. 

\begin{figure}[p]
\vfill
\setlength{\unitlength}{1.0cm}
\begin{picture} (18.0,15.0)
\put (1.0,0.0){\epsfig{figure=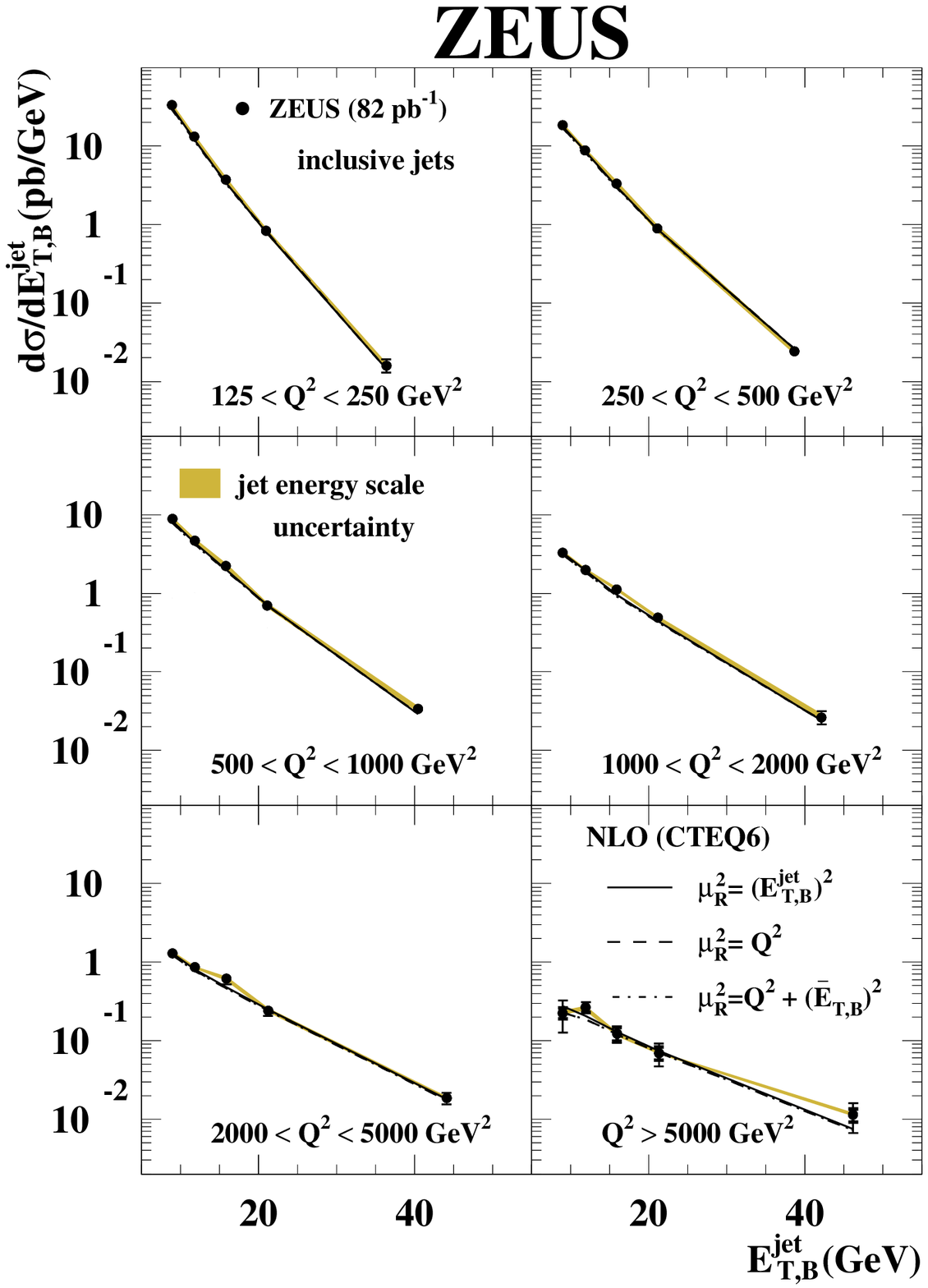,width=15cm}}
\end{picture}
\caption
{\it 
The measured differential cross-section $\setjb$ for inclusive-jet
production with $\etjb >$~8~{\rm GeV} and $-2<\etajb<1.5$ 
in different regions of $\q2$ (dots).
The NLO QCD calculations with $\mu_R^2=(\etjb)^2$ (solid lines),
$\mu_R^2=Q^2$ (dashed lines) and $\mu_R^2=\q2+\overline{E}_{T,{\rm B}}^2$ (dash-dotted lines), 
corrected for hadronisation and $\z0$ effects
and using the CTEQ6 parameterisations of the proton PDFs, are also
shown. Other details as in the caption to Fig.~\ref{fig110}.
}
\label{fig8}
\vfill
\end{figure}

\begin{figure}[p]
\vfill
\setlength{\unitlength}{1.0cm}
\begin{picture} (18.0,15.0)
\put (1.0,0.0){\epsfig{figure=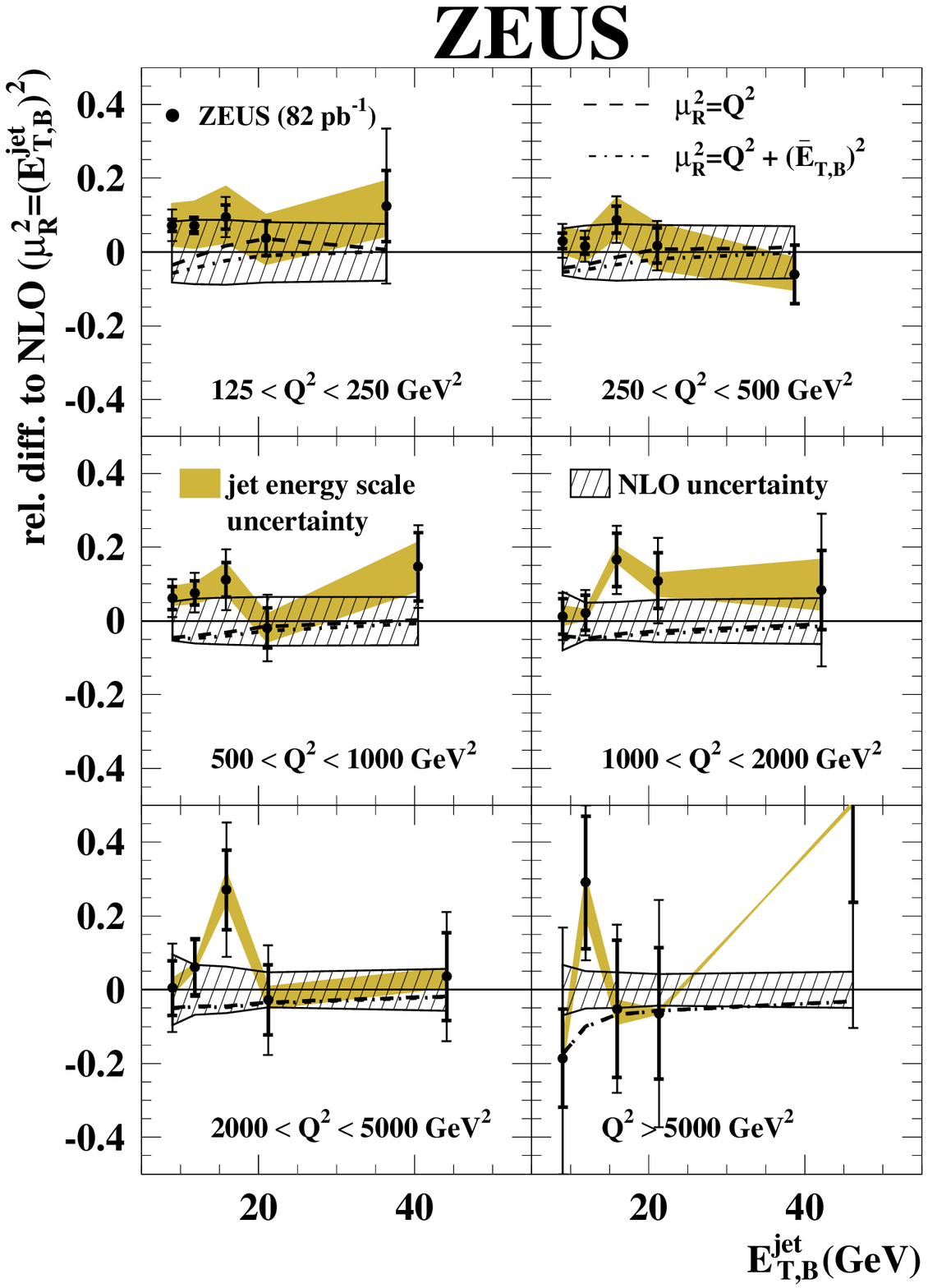,width=15cm}}
\end{picture}
\caption
{\it 
Relative differences between the measured differential cross-sections
$\setjb$ presented in Fig.~\ref{fig8} and the NLO QCD calculations
with $\mu_R^2=(E_{T,B}^{jet})^2$ (dots). 
The relative differences between the NLO QCD calculations with $\mu_R^2=\q2$
($\mu_R^2=\q2+\overline{E}_{T,{\rm B}}^2$) and those with
$\mu_R^2=(E_{T,B}^{jet})^2$
are also shown as dashed (dash-dotted) lines.
Other details as in the caption to Fig.~\ref{fig110}.
}
\label{fig9}
\vfill
\end{figure}

Inclusive-jet cross sections were measured in the kinematic region
$\q2>125$~\g2\ and $|\cgh|<0.65$. These cross sections include every
jet of hadrons in the event with $\etjb>8$~GeV and $-2<\etajb<1.5$ and
were corrected for detector and QED radiative effects. The cross
sections for different regions of $\q2$ as a function of $\etjb$ are
presented in Fig.~\ref{fig8} and Table~\ref{inclusive_jets}. 
The measured cross sections exhibit a steep fall-off within the $\etjb$ 
range considered. As $\q2$ increases, the $\etjb$ dependence of the 
cross section becomes less steep. 

The NLO QCD predictions with $\mu_R^2=(\etjb)^2$ are compared to the
measurements in Fig.~\ref{fig8}. They give a good description of the
data. To study the scale dependence, calculations using $\mu_R^2=Q^2$ or 
$\mu_R^2=\q2+\overline{E}_{T,{\rm B}}^2$ are
also compared to the data in Fig.~\ref{fig8}; they provide an approximately
equally good description of the data.

Figure~\ref{fig9} shows the relative difference of the measured
differential cross sections to the NLO QCD calculations with
$\mu_R^2=(E^{\rm jet}_{T,{\rm B}})^2$. 
The uncertainty of the NLO QCD calculations and the results of
the theory calculations using either $\mu_R^2=\q2$ or
$\mu_R^2=\q2+\overline{E}_{T,{\rm B}}^2$ are also shown. 
The data are well described by the
predictions.

As indicated in Section~\ref{nlo}, the theoretical uncertainties in both
the inclusive-jet and dijet regimes are dominated by the contribution from
higher 
orders as estimated by varying the renormalisation scale. 
As was shown in Figs.~\ref{fig11} and~\ref{fig11a},
the scale uncertainty decreases as $Q^2$ increases; 
it is also lower for high values of $\xi$ than for low values of this
variable. The contribution from the PDF uncertainty is approximately constant
in all variables and non-negligible. 
Especially at high values of $\xi$ ($\etjb$) for
the dijet (inclusive-jet) analysis, the PDF uncertainty is even dominant.
In these regions, in which the gluon-induced contribution is still 
substantial (see Figs.~\ref{gluefrac1} to~\ref{gluefrac}), 
the data will be able to  
constrain further the gluon density in the proton.  

\section{Summary and conclusions}

Inclusive-jet and dijet cross sections have been measured at high $\q2$
values at HERA for $\sqrt s=318$~GeV. The jets were reconstructed using
the $\kt$ cluster algorithm in the longitudinally invariant inclusive
mode in the Breit frame. The dijet cross sections were measured as
functions of $\q2$, $x_{\rm Bj}$, $\overline{E}_T$, $M_{\rm jj}$, 
$\eta^{\prime}$ and $\xi$. In
addition, inclusive-jet and dijet measurements were performed in different 
regions of $\q2$ as functions of $\etjb$ and $\xi$, respectively. 
The data are well
described by the NLO QCD predictions. The cross sections in different
regions of $\q2$ are shown to be sensitive to the gluon
density of the proton.
The precise measurements obtained here
are therefore of particular relevance for improving the determination of the
gluon density in future QCD fits.

\vspace{0.5cm}
\noindent {\Large\bf Acknowledgements}
\vspace{0.3cm}

We thank the DESY Directorate for their strong support and encouragement.
The remarkable achievements of the HERA machine group were essential for
the successful completion of this work and are greatly appreciated. We are
grateful for the support of the DESY computing and network services. The
design, construction and installation of the ZEUS detector have been made
possible owing to the ingenuity and effort of many people from DESY and home
institutes who are not listed as authors.

\newpage

\providecommand{\etal}{et al.\xspace}
\providecommand{\coll}{Coll.\xspace}
\catcode`\@=11
\def\@bibitem#1{%
\ifmc@bstsupport
  \mc@iftail{#1}%
    {;\newline\ignorespaces}%
    {\ifmc@first\else.\fi\orig@bibitem{#1}}
  \mc@firstfalse
\else
  \mc@iftail{#1}%
    {\ignorespaces}%
    {\orig@bibitem{#1}}%
\fi}%
\catcode`\@=12

\newpage

 \begin{table}[htbp]
  \begin{center}
    \begin{tabular}{||c|cccc||c||c||}
      \hline 
$Q^{2}$ bin & $d\sigma/dQ^{2}$ &
 & & & & \\ 
(${\rm GeV^2}$) & ${\rm (pb/GeV^{2})}$ &
$\delta_{\rm stat}$ & $\delta_{\rm syst}$ & $\delta_{\rm ES}$ & $C_{\rm QED}$ & $C_{\rm had}$ \\ 
      \hline 
      \hline 
125-250&$0.1183$&$\pm 0.0033$&$^{+ 0.0041}_{- 0.0040}$&$^{+ 0.0081}_{- 0.0077}$&$ 0.96$&$ 0.85$\\ 
250-500&$0.0589$&$\pm 0.0018$&$^{+ 0.0025}_{- 0.0023}$&$^{+ 0.0032}_{- 0.0030}$&$ 0.94$&$ 0.91$\\ 
500-1000&$0.02061$&$\pm 0.00074$&$^{+ 0.00094}_{- 0.00094}$&$^{+ 0.00096}_{- 0.00083}$&$ 0.93$&$ 0.92$\\ 
1000-2000&$0.00602$&$\pm 0.00028$&$^{+ 0.00018}_{- 0.00018}$&$^{+ 0.00014}_{- 0.00013}$&$ 0.91$&$ 0.95$\\ 
2000-5000&$0.001189$&$\pm 0.000075$&$^{+ 0.000083}_{- 0.000083}$&$^{+ 0.000025}_{- 0.000020}$&$ 0.96$&$ 0.96$\\ 
      \hline 
    \end{tabular}
    \caption{Dijet cross-section $d\sigma/dQ^2$ for jets
    of hadrons in the Breit frame selected with the longitudinally invariant
    $k_T$ cluster algorithm. The statistical, uncorrelated systematic and
    jet-energy-scale ({\rm ES}) uncertainties are shown separately. 
    The multiplicative corrections applied to the data to correct for QED
    radiative effects, $C_{\rm QED}$, and the corrections for hadronisation
    effects to be applied to the parton-level NLO QCD calculations, 
    $C_{\rm had}$, are shown in the last two columns.}
    \label{q2_inc_dijet_tab}
  \end{center}
\end{table}

 \begin{table}[htbp]
  \begin{center}
    \begin{tabular}{||c|cccc||c||c||}
      \hline 
$x_{{\rm Bj}}$ bin & $d\sigma/dx_{{\rm Bj}}$ &
& & & &  \\ 
 & (pb) &
$\delta_{\rm stat}$ & $\delta_{\rm syst}$ & $\delta_{\rm ES}$ & $C_{\rm QED}$ & $C_{\rm had}$ \\ 
      \hline 
      \hline 
0.0001-0.01&$1599$&$\pm 45$&$^{+ 54}_{- 52}$&$^{+ 101}_{- 92}$&$ 0.95$&$ 0.86$\\ 
0.01-0.02&$1849$&$\pm 48$&$^{+ 37}_{- 37}$&$^{+ 98}_{- 95}$&$ 0.94$&$ 0.93$\\ 
0.02-0.035&$647$&$\pm 24$&$^{+ 26}_{- 25}$&$^{+ 26}_{- 23}$&$ 0.92$&$ 0.92$\\ 
0.035-0.07&$141.7$&$\pm 7.5$&$^{+ 4.0}_{- 4.0}$&$^{+ 2.9}_{- 2.6}$&$ 0.95$&$ 0.91$\\ 
0.07-0.1&$21.1$&$\pm 3.2$&$^{+ 3.3}_{- 3.3}$&$^{+ 0.5}_{- 0.1}$&$ 0.89$&$ 0.90$\\ 
      \hline 
    \end{tabular}
    \caption{Dijet cross-section $d\sigma/dx_{{\rm Bj}}$ for jets
    of hadrons in the Breit frame selected with the longitudinally invariant
    $k_T$ cluster algorithm. Other details as in the caption to
    Table~\ref{q2_inc_dijet_tab}.}
    \label{xBj_inc_dijet_tab}
  \end{center}
\end{table}

\begin{table}[htbp]
  \begin{center}
    \begin{tabular}{||c|cccc||c||c||}
      \hline 
${\overline{E}_T}$ bin & $d\sigma/d{\overline{E}_T}$ &
 & & & & \\ 
(GeV) & (pb/GeV) &
$\delta_{\rm stat}$ & $\delta_{\rm syst}$ & $\delta_{\rm ES}$ & $C_{\rm QED}$
      & $C_{\rm had}$ \\ 
      \hline 
      \hline 
10-16&$5.71$&$\pm 0.11$&$^{+ 0.12}_{- 0.12}$&$^{+ 0.30}_{- 0.26}$&$ 0.94$&$ 0.90$\\ 
16-22&$1.935$&$\pm 0.064$&$^{+ 0.086}_{- 0.084}$&$^{+ 0.081}_{- 0.091}$&$ 0.94$&$ 0.90$\\ 
22-30&$0.361$&$\pm 0.023$&$^{+ 0.023}_{- 0.023}$&$^{+ 0.017}_{- 0.016}$&$ 0.94$&$ 0.91$\\ 
30-60&$0.0371$&$\pm 0.0043$&$^{+ 0.0076}_{- 0.0076}$&$^{+ 0.0021}_{- 0.0018}$&$ 0.97$&$ 0.91$\\ 
      \hline 
    \end{tabular}
    \caption{Dijet cross-section $d\sigma/d{\overline{E}_T}$ for jets
    of hadrons in the Breit frame selected with the longitudinally invariant
    $k_T$ cluster algorithm. Other details as in the caption to
    Table~\ref{q2_inc_dijet_tab}.}
    \label{ET_inc_dijet_tab}
  \end{center}
\end{table}

 \begin{table}[htbp]
  \begin{center}
    \begin{tabular}{||c|cccc||c||c||}
      \hline 
$M_{{\rm jj}}$ bin & $d\sigma/dM_{{\rm jj}}$ &
& & & & \\ 
(GeV) & (pb/GeV) &
$\delta_{\rm stat}$ & $\delta_{\rm syst}$ & $\delta_{\rm ES}$ & $C_{\rm QED}$ & $C_{\rm had}$ \\ 
      \hline 
      \hline 
20-32&$2.382$&$\pm 0.051$&$^{+ 0.055}_{- 0.054}$&$^{+ 0.123}_{- 0.109}$&$ 0.95$&$ 0.91$\\ 
32-45&$1.134$&$\pm 0.034$&$^{+ 0.047}_{- 0.047}$&$^{+ 0.049}_{- 0.055}$&$ 0.94$&$ 0.89$\\ 
45-65&$0.222$&$\pm 0.012$&$^{+ 0.006}_{- 0.006}$&$^{+ 0.009}_{- 0.009}$&$ 0.92$&$ 0.89$\\ 
65-120&$0.0180$&$\pm 0.0021$&$^{+ 0.0017}_{- 0.0017}$&$^{+ 0.0013}_{- 0.0009}$&$ 0.97$&$ 0.94$\\ 
      \hline 
    \end{tabular}
    \caption{Dijet cross-section $d\sigma/dM_{{\rm jj}}$ for jets
    of hadrons in the Breit frame selected with the longitudinally invariant
    $k_T$ cluster algorithm. Other details as in the caption to Table~\ref{q2_inc_dijet_tab}.}
    \label{Mjj_inc_dijet_tab}
  \end{center}
\end{table}

 \begin{table}[htbp]
  \begin{center}
    \begin{tabular}{||c|cccc||c||c||}
      \hline 
$\eta^{'}$ bin & $d\sigma/d\eta{'}$ &
& & & & \\ 
 & (pb) &
$\delta_{\rm stat}$ & $\delta_{\rm syst}$ & $\delta_{\rm ES}$ & $C_{\rm QED}$ & $C_{\rm had}$ \\ 
      \hline 
      \hline 
0-0.1&$87.9$&$\pm 3.4$&$^{+ 4.0}_{- 3.9}$&$^{+ 3.7}_{- 3.7}$&$ 0.95$&$ 0.91$\\ 
0.1-0.25&$87.0$&$\pm 2.8$&$^{+ 1.6}_{- 1.6}$&$^{+ 3.6}_{- 3.7}$&$ 0.94$&$ 0.94$\\ 
0.25-0.45&$68.5$&$\pm 2.1$&$^{+ 1.3}_{- 1.1}$&$^{+ 3.5}_{- 3.0}$&$ 0.95$&$ 0.91$\\ 
0.45-0.65&$42.7$&$\pm 1.6$&$^{+ 1.0}_{- 1.0}$&$^{+ 2.5}_{- 2.2}$&$ 0.94$&$ 0.87$\\ 
0.65-1.60&$6.07$&$\pm 0.28$&$^{+ 0.54}_{- 0.52}$&$^{+ 0.37}_{- 0.34}$&$ 0.91$&$ 0.86$\\ 
      \hline 
    \end{tabular}
    \caption{Dijet cross-section $d\sigma/d\eta^{'}$ for jets
    of hadrons in the Breit frame selected with the longitudinally invariant
    $k_T$ cluster algorithm. Other details as in the caption to Table~\ref{q2_inc_dijet_tab}.}
    \label{eta_inc_dijet_tab}
  \end{center}
\end{table}

 \begin{table}[htbp]
  \begin{center}
    \begin{tabular}{||c|cccc||c||c||}
      \hline 
$\log_{10}\xi$ bin & $d\sigma/d\log_{10}\xi$ &
& & & & \\ 
 & (pb) &
$\delta_{\rm stat}$ & $\delta_{\rm syst}$ & $\delta_{\rm ES}$ & $C_{\rm QED}$ & $C_{\rm had}$ \\ 
      \hline 
      \hline 
--2 - --1.5&$22.22$&$\pm 0.78$&$^{+ 0.72}_{- 0.71}$&$^{+ 1.34}_{- 1.17}$&$ 0.95$&$ 0.88$\\ 
--1.5 - --1.35&$74.4$&$\pm 2.6$&$^{+ 2.8}_{- 2.8}$&$^{+ 3.2}_{- 3.4}$&$ 0.94$&$ 0.92$\\ 
--1.35 - --1.1&$73.9$&$\pm 1.9$&$^{+ 2.2}_{- 2.0}$&$^{+ 3.4}_{- 3.2}$&$ 0.94$&$ 0.92$\\ 
--1.1 - --0.85&$31.0$&$\pm 1.3$&$^{+ 2.4}_{- 2.4}$&$^{+ 1.5}_{- 1.4}$&$ 0.94$&$ 0.87$\\ 
--0.85 - --0.5&$3.93$&$\pm 0.36$&$^{+ 0.48}_{- 0.48}$&$^{+ 0.32}_{- 0.21}$&$ 0.93$&$ 0.79$\\ 
      \hline 
    \end{tabular}
    \caption{Dijet cross-section $d\sigma/d\log_{10}\xi$ for jets
    of hadrons in the Breit frame selected with the longitudinally invariant
    $k_T$ cluster algorithm. Other details as in the caption to Table~\ref{q2_inc_dijet_tab}.}
    \label{xi_inc_dijet_tab}
  \end{center}
\end{table}

 \begin{table}[htbp]
  \begin{center}
    \begin{tabular}{||c|cccc||c||c||}
      \hline 
$\log_{10}\xi$ bin & $d\sigma/d\log_{10}\xi$ &
& & & &  \\ 
& (pb) &
$\delta_{\rm stat}$ & $\delta_{\rm syst}$ & $\delta_{\rm ES}$ & $C_{\rm QED}$ & $C_{\rm had}$ \\ 
      \hline 
      \hline 
      \multicolumn{7}{||c||}{125~$< Q^2 <$~250~${\rm GeV^2}$ } \\
      \hline 
      --2.00 - --1.50&$9.17$&$\pm 0.48$&$^{+ 0.44}_{- 0.44}$&$^{+ 0.70}_{- 0.62}$&$ 0.97$&$ 0.83$\\ 
      --1.50 - --1.35&$25.0$&$\pm 1.4$&$^{+ 1.8}_{- 1.8}$&$^{+ 1.4}_{- 0.7}$&$ 0.95$&$ 0.89$\\ 
      --1.35 - --1.10&$19.74$&$\pm 0.93$&$^{+ 0.43}_{- 0.41}$&$^{+ 1.17}_{- 1.65}$&$ 0.95$&$ 0.86$\\ 
      --1.10 - --0.50&$2.52$&$\pm 0.20$&$^{+ 0.10}_{- 0.10}$&$^{+ 0.26}_{- 0.19}$&$ 0.99$&$ 0.83$\\ 
      \hline 
      \multicolumn{7}{||c||}{250~$< Q^2 <$~500~${\rm GeV^2}$ } \\
      \hline 
      --2.00 - --1.50&$8.65$&$\pm 0.49$&$^{+ 0.59}_{- 0.57}$&$^{+ 0.42}_{- 0.38}$&$ 0.94$&$ 0.93$\\ 
      --1.50 - --1.30&$23.4$&$\pm 1.2$&$^{+ 1.5}_{- 1.4}$&$^{+ 1.1}_{- 1.3}$&$ 0.94$&$ 0.93$\\ 
      --1.30 - --1.00&$16.18$&$\pm 0.83$&$^{+ 0.55}_{- 0.44}$&$^{+ 1.02}_{- 0.78}$&$ 0.93$&$ 0.89$\\ 
      --1.00 - --0.50&$1.77$&$\pm 0.21$&$^{+ 0.09}_{- 0.09}$&$^{+ 0.10}_{- 0.11}$&$ 0.94$&$ 0.88$\\ 
      \hline 
      \multicolumn{7}{||c||}{500~$< Q^2 <$~1000~${\rm GeV^2}$ } \\
      \hline 
      --1.90 - --1.50&$4.32$&$\pm 0.40$&$^{+ 0.15}_{- 0.18}$&$^{+ 0.23}_{- 0.17}$&$ 0.94$&$ 0.92$\\ 
      --1.50 - --1.20&$17.50$&$\pm 0.88$&$^{+ 1.00}_{- 1.00}$&$^{+ 0.72}_{- 0.75}$&$ 0.94$&$ 0.94$\\ 
      --1.20 - --0.90&$9.45$&$\pm 0.63$&$^{+ 0.90}_{- 0.90}$&$^{+ 0.50}_{- 0.35}$&$ 0.91$&$ 0.92$\\ 
      --0.90 - --0.60&$1.48$&$\pm 0.26$&$^{+ 0.09}_{- 0.09}$&$^{+ 0.06}_{- 0.06}$&$ 0.98$&$ 0.86$\\ 
      \hline 
      \multicolumn{7}{||c||}{1000~$< Q^2 <$~2000~${\rm GeV^2}$ } \\
      \hline 
      --1.70 - --1.40&$2.72$&$\pm 0.37$&$^{+ 0.33}_{- 0.33}$&$^{+ 0.06}_{- 0.12}$&$ 0.93$&$ 0.94$\\ 
      --1.40 - --1.20&$10.98$&$\pm 0.86$&$^{+ 1.23}_{- 1.23}$&$^{+ 0.22}_{- 0.01}$&$ 0.94$&$ 0.98$\\ 
      --1.20 - --1.00&$10.37$&$\pm 0.84$&$^{+ 0.48}_{- 0.48}$&$^{+ 0.26}_{- 0.28}$&$ 0.92$&$ 0.96$\\ 
      --1.00 - --0.60&$2.17$&$\pm 0.25$&$^{+ 0.15}_{- 0.15}$&$^{+ 0.06}_{- 0.08}$&$ 0.85$&$ 0.89$\\ 
      \hline 
      \multicolumn{7}{||c||}{2000~$< Q^2 <$~5000~${\rm GeV^2}$ } \\
      \hline 
      --1.60 - --1.20&$2.14$&$\pm 0.26$&$^{+ 0.14}_{- 0.14}$&$^{+ 0.01}_{- 0.02}$&$ 0.92$&$ 0.99$\\ 
      --1.20 - --1.00&$8.28$&$\pm 0.80$&$^{+ 0.54}_{- 0.54}$&$^{+ 0.14}_{- 0.18}$&$ 1.00$&$ 0.96$\\ 
      --1.00 - --0.60&$2.48$&$\pm 0.30$&$^{+ 0.16}_{- 0.16}$&$^{+ 0.08}_{- 0.04}$&$ 0.93$&$ 0.94$\\ 
      \hline 
    \end{tabular}
    \caption{Dijet cross-sections $d\sigma/d\log_{10}\xi$ for jets
    of hadrons in the Breit frame selected with the longitudinally invariant
    $k_T$ cluster algorithm in different regions of $Q^2$. 
    Other details as in the caption to
    Table~\ref{q2_inc_dijet_tab}.}
    \label{dd_xi_inc_dijet_tab}
  \end{center}
\end{table}
 
\renewcommand{\arraystretch}{0.95}

\begin{table}
\vspace*{-2.cm}
\begin{center}
    \begin{tabular}{||c|cccc||c||c||}
\hline
  $\etjb$ bin
& $d\sigma/d\etjb$
&
&
&
&
& \\
  (GeV)
& (pb/GeV)
& $\delta_{\rm stat}$
& $\delta_{\rm syst}$
& $\delta_{\rm ES}$
& $C_{\rm QED}$
& $C_{\rm had}$\\
\hline
\multicolumn{7}{||c||}{$125 < Q^2 < 250$~GeV$^2$} \\
\hline
 8-10
&32.97
&0.49
&{\small ${}_{-1.21}^{+1.21}$}
&{\small ${}_{-1.69}^{+1.81}$}
&0.96
&0.90
\\
 10-14
&13.00
&0.22
&{\small ${}_{-0.19}^{+0.19}$}
&{\small ${}_{-0.75}^{+0.79}$}
&0.98
&0.94
\\
 14-18
&3.71
&0.11
&{\small ${}_{-0.15}^{+0.15}$}
&{\small ${}_{-0.24}^{+0.28}$}
&0.97
&0.94
\\
 18-25
&0.835
&0.037
&{\small ${}_{-0.012}^{+0.013}$}
&{\small ${}_{-0.056}^{+0.051}$}
&0.94
&0.93
\\
 25-100
&0.0160
&0.0014
&{\small ${}_{-0.0027}^{+0.0027}$}
&{\small ${}_{-0.0011}^{+0.0010}$}
&0.97
&0.86
\\
\hline
\multicolumn{7}{||c||}{$250 < Q^2 < 500$~GeV$^2$} \\
\hline
 8-10
&18.40
&0.38
&{\small ${}_{-0.74}^{+0.74}$}
&{\small ${}_{-0.60}^{+0.68}$}
&0.94
&0.92
\\
 10-14
&8.74
&0.19
&{\small ${}_{-0.30}^{+0.30}$}
&{\small ${}_{-0.35}^{+0.33}$}
&0.96
&0.95
\\
 14-18
&3.30
&0.11
&{\small ${}_{-0.15}^{+0.15}$}
&{\small ${}_{-0.14}^{+0.18}$}
&0.96
&0.97
\\
 18-25
&0.889
&0.042
&{\small ${}_{-0.041}^{+0.041}$}
&{\small ${}_{-0.057}^{+0.052}$}
&0.92
&0.97
\\
 25-100
&0.0242
&0.0020
&{\small ${}_{-0.0005}^{+0.0005}$}
&{\small ${}_{-0.0011}^{+0.0012}$}
&0.95
&0.91
\\
\hline
\multicolumn{7}{||c||}{$500 < Q^2 < 1000$~GeV$^2$} \\
\hline
 8-10
&8.79
&0.26
&{\small ${}_{-0.34}^{+0.34}$}
&{\small ${}_{-0.15}^{+0.26}$}
&0.96
&0.91
\\
 10-14
&4.69
&0.14
&{\small ${}_{-0.19}^{+0.19}$}
&{\small ${}_{-0.13}^{+0.11}$}
&0.94
&0.95
\\
 14-18
&2.239
&0.093
&{\small ${}_{-0.137}^{+0.137}$}
&{\small ${}_{-0.074}^{+0.091}$}
&0.93
&0.98
\\
 18-25
&0.701
&0.039
&{\small ${}_{-0.051}^{+0.051}$}
&{\small ${}_{-0.026}^{+0.026}$}
&0.96
&0.99
\\
 25-100
&0.0335
&0.0027
&{\small ${}_{-0.0018}^{+0.0018}$}
&{\small ${}_{-0.0019}^{+0.0019}$}
&0.96
&0.97
\\
\hline
\multicolumn{7}{||c||}{$1000 < Q^2 < 2000$~GeV$^2$} \\
\hline
 8-10
&3.30
&0.16
&{\small ${}_{-0.14}^{+0.14}$}
&{\small ${}_{-0.08}^{+0.09}$}
&0.93
&0.93
\\
 10-14
&1.985
&0.091
&{\small ${}_{-0.077}^{+0.077}$}
&{\small ${}_{-0.037}^{+0.017}$}
&0.91
&0.95
\\
 14-18
&1.115
&0.069
&{\small ${}_{-0.056}^{+0.056}$}
&{\small ${}_{-0.008}^{+0.034}$}
&0.98
&0.99
\\
 18-25
&0.492
&0.034
&{\small ${}_{-0.039}^{+0.039}$}
&{\small ${}_{-0.018}^{+0.009}$}
&0.93
&0.99
\\
 25-100
&0.0263
&0.0026
&{\small ${}_{-0.0043}^{+0.0043}$}
&{\small ${}_{-0.0013}^{+0.0020}$}
&1.00
&1.00
\\
\hline
\multicolumn{7}{||c||}{$2000 < Q^2 < 5000$~GeV$^2$} \\
\hline
 8-10
&1.292
&0.095
&{\small ${}_{-0.120}^{+0.120}$}
&{\small ${}_{-0.022}^{+0.033}$}
&0.92
&0.90
\\
 10-14
&0.858
&0.060
&{\small ${}_{-0.024}^{+0.024}$}
&{\small ${}_{-0.006}^{+0.007}$}
&0.90
&0.93
\\
 14-18
&0.612
&0.052
&{\small ${}_{-0.070}^{+0.070}$}
&{\small ${}_{-0.017}^{+0.023}$}
&1.02
&1.00
\\
 18-25
&0.242
&0.024
&{\small ${}_{-0.028}^{+0.028}$}
&{\small ${}_{-0.006}^{+0.009}$}
&0.96
&1.00
\\
 25-100
&0.0185
&0.0021
&{\small ${}_{-0.0023}^{+0.0023}$}
&{\small ${}_{-0.0005}^{+0.0004}$}
&0.91
&0.99
\\
\hline
\multicolumn{7}{||c||}{$5000 < Q^2 < 100000$~GeV$^2$} \\
\hline
 8-10
&0.225
&0.037
&{\small ${}_{-0.091}^{+0.091}$}
&{\small ${}_{-0.006}^{+0.011}$}
&0.99
&0.93
\\
 10-14
&0.267
&0.037
&{\small ${}_{-0.023}^{+0.022}$}
&{\small ${}_{-0.019}^{+0.003}$}
&0.96
&0.93
\\
 14-18
&0.122
&0.024
&{\small ${}_{-0.017}^{+0.017}$}
&{\small ${}_{-0.005}^{+0.003}$}
&0.97
&0.98
\\
 18-25
&0.070
&0.013
&{\small ${}_{-0.019}^{+0.019}$}
&{\small ${}_{-0.000}^{+0.001}$}
&0.98
&0.99
\\
 25-100
&0.0114
&0.0022
&{\small ${}_{-0.0042}^{+0.0042}$}
&{\small ${}_{-0.0001}^{+0.0003}$}
&0.99
&1.00
\\
\hline
    \end{tabular}
 \caption{
     Inclusive jet cross-sections $d\sigma/d\etjb$ for jets
    of hadrons in the Breit frame selected with the longitudinally invariant
    $k_T$ cluster algorithm in different regions of $Q^2$. 
    Other details as in the caption to
    Table~\ref{q2_inc_dijet_tab}.}
 \label{inclusive_jets}
\end{center}
\end{table}

\newpage
\clearpage

\newpage
\clearpage

\newpage
\clearpage

\newpage
\clearpage

\newpage
\clearpage

\newpage
\clearpage

\newpage
\clearpage

\newpage
\clearpage

\newpage
\clearpage

\newpage
\clearpage

\newpage
\clearpage

\newpage
\clearpage

\end{document}